\documentclass[12pt]{amsart}
\usepackage{amsmath}
\usepackage{amsxtra}
\usepackage{amscd}
\usepackage{amsthm}
\usepackage{amsfonts}
\usepackage{amssymb}
\usepackage{eucal}
\def\({\left(}
\def\){\right)}
\newcommand{\ds}[1]{\displaystyle #1}
\renewcommand{\Re}{\mathop{\rm Re}}  

\newcommand{\nn}{\nonumber}
\newcommand{\bea}{\begin{eqnarray}}
\newcommand{\ena}{\end{eqnarray}}
\newcommand{\be}{\begin{eqnarray*}}
\newcommand{\en}{\end{eqnarray*}}
\newcommand{\ba}{\begin{array}}
\newcommand{\ea}{\end{array}}

\newcommand{\R}{{\mathbb R}}
\newcommand{\C}{{\mathbb C}}
\newcommand{\Z}{{\mathbb Z}} 
\newcommand{\Q}{{\mathbb Q}} 
\newcommand{\cP}{\mathcal{P}}

\newcommand{\Rc}{\check{R}}

\newcommand{\slt}{\mathfrak{sl}_2}
\newcommand{\res}{{\rm res}}
\newcommand{\id}{{\rm id}}
\newcommand{\tr}{{\rm tr}}
\newcommand{\wt}{{\rm wt}\,}
\newcommand{\Tr}{{\rm Tr}}
\newcommand{\vac}{{\rm vac}}

\newcommand{\End}{\mathop{\rm End}}
\newcommand{\Hom}{\mathop{\rm Hom}}
\newcommand{\Ker}{\mathop{\rm Ker}}
\renewcommand{\Im}{\mathop{\rm Im}}
\newenvironment{tenumerate}{
  \begin{enumerate}
  
  }{\end{enumerate}}
\newcommand{\bi}{\begin{tenumerate}}
\newcommand{\ei}{\end{tenumerate}}
\newcommand{\isoto}[1][]%
{{\mathop{\buildrel{\sim}\over\longrightarrow}\limits_{#1}}}

\newcommand{\la}{\lambda}

\newcommand{\e}{\epsilon}
\newcommand{\eb}{\bar{\epsilon}}
\newcommand{\z}{\zeta}
\newcommand{\bs}{\mathbf{s}}
\newcommand{\ot}{\tilde{\omega}}
\newcommand{\hti}{h^{Corr}}
\newcommand{\pt}{\tilde{p}}
\newcommand{\rt}{\tilde{\rho}}
\newcommand{\Ft}{\tilde{G}}
\newcommand{\F}{{G}}

\numberwithin{equation}{section}
\newtheorem{thm}{Theorem}[section]
\newtheorem{prop}[thm]{Proposition}
\newtheorem{lem}[thm]{Lemma}

\begin{document} 
\title[Reduced qKZ equation]
{Reduced qKZ equation and correlation
functions of the XXZ model}
\date{\today}
\author{H.~Boos, M.~Jimbo, T.~Miwa, F.~Smirnov and Y.~Takeyama}
\address{HB: Physics Department, University of Wuppertal, D-42097,
Wuppertal, Germany\footnote{
on leave of absence from the Institute for High Energy Physics, Protvino, 
142281, Russia}}\email{boos@physik.uni-wuppertal.de}
\address{MJ: Graduate School of Mathematical Sciences, The
University of Tokyo, Tokyo 153-8914, Japan}\email{jimbomic@ms.u-tokyo.ac.jp}
\address{TM: Department of Mathematics, Graduate School of Science,
Kyoto University, Kyoto 606-8502, 
Japan}\email{tetsuji@math.kyoto-u.ac.jp}
\address{FS\footnote{Membre du CNRS}: Laboratoire de Physique Th{\'e}orique et
Hautes Energies, Universit{\'e} Pierre et Marie Curie,
Tour 16 1$^{\rm er}$ {\'e}tage, 4 Place Jussieu
75252 Paris Cedex 05, France}\email{smirnov@lpthe.jussieu.fr}
\address{YT: Graduate School of Pure and Applied Sciences, 
Tsukuba University, Tsukuba, Ibaraki 305-8571, Japan}
\email{takeyama@math.tsukuba.ac.jp}

\begin{abstract}
Correlation functions of the XXZ model
in the massive and massless regimes  
are known to satisfy a system of linear equations. 
The main relations among them are the difference equations 
obtained from the qKZ equation by specializing the 
variables $(\la_1,\cdots,\la_{2n})$ as 
$(\la_1,\cdots,\la_n,\la_n+1,\cdots,\la_1+1)$. 
We call it the reduced qKZ equation.  
In this article we construct a special family of 
solutions to this system. They can be written  
as linear combinations of 
products of two transcendental functions 
$\ot,\omega$ with coefficients being rational functions. 
We show that correlation functions of the XXZ model 
in the massive regime are given by these formulas 
with an appropriate choice of $\ot,\omega$. 
We also present a conjectural formula in the massless regime. 
\end{abstract}

\maketitle

\setcounter{section}{0}
\setcounter{equation}{0}

\section{Introduction}\label{sec:1}

Correlation functions of integrable spin chains in one dimension 
have been a subject of intensive study. 
In the most typical case of the XXX and the XXZ models, 
they are given in the form 
\bea
g_{2n}(\la_1,\cdots,\la_n,\la_n+1,\cdots,\la_1+1),
\label{eq:reduced}
\ena
where $g_{2n}(\la_1,\cdots,\la_{2n})$ is a certain solution of 
the quantum Knizhnik-Zamolodchikov (qKZ) equation 
of 
`level $-4$' \cite{IIJMNT}.  
The specialized function \eqref{eq:reduced} satisfies
a similar system of difference equations, 
which we refer to as the reduced qKZ equation. 
Multi-dimensional integral representations for 
the correlation functions have been known for some time \cite{JMMN,JM,KMT}. 
It has been recognized through the recent works 
\cite{BK, TKS,KSTS1,KSTS2} that, 
at least for small $n$, 
these multiple integrals can be reduced to 
one-dimensional integrals, thus 
making it possible to evaluate them explicitly.  
The reason for this mysterious reducibility was 
subsequently explained 
via a duality between the qKZ equation of level 
$0$ and level $-4$ \cite{BKS1,BKS2,BKS3}. 
In these works, the following general 
{\it Ansatz} was proposed. 
Consider an inhomogeneous XXX model where 
each site $j$ carries an independent spectral parameter $\la_j$.
Then the ground state average of products of elementary 
operators $(E_{\e_j,\eb_j})_j$ can be written in the form 
\bea
&&
\langle {\rm vac}|
(E_{\e_1,\eb_1})_1\cdots (E_{\e_n,\eb_n})_n
|{\rm vac} \rangle
\label{eq:XXZ-Ansatz}\\&&\qquad =\sum_{m=0}^{[n/2]}\sum_{I,J}
\prod_{p=1}^m \omega^{XXX}(\la_{i_p}-\la_{j_p})\, 
f_{n,I,J}(\la_1,\cdots,\la_n),   
\nn
\ena
where $\omega^{XXX}(\la)$ is a certain transcendental function,  
$f_{n,I,J}(\la _1,\cdots ,\la _n)$ 
are rational functions, 
and the sum is taken over 
$I=(i_1,\cdots,i_m),J=(j_1,\cdots,j_m)\in\{1,\ldots,n\}^m$, 
such that $i_1,\ldots,i_m,j_1,\ldots,j_m$ are distinct,
$i_p<j_p$ $(1\leq p\leq m)$ and $i_1<\cdots<i_m$.
However the rational functions $f_{n,I,J}$ were not described. 
In \cite{BKS4}, it was explained 
that a similar formula continues to hold for 
the correlation functions of quantum group reduction of the XXZ model. 



In the present paper we consider the genuine XXZ model (without quantum group reduction).
Let $\Delta=\cos\pi\nu$ be the coupling constant 
of the XXZ chain (see \eqref{eq:XXZ}), 
and set $q=e^{\pi i\nu}$.   
We arrange the correlation functions 
into a function $h^{\it Corr}_n$ with values in $V^{\otimes 2n}$,  
where $V=\C v_+\oplus \C v_-$. 
Set $s=v_+\otimes v_--v_-\otimes v_+$, 
${\bf s}_n=\prod _{p=1}^n s_{p,2n-p+1}\in V^{\otimes 2n}$    
(as usual, the lower indices indicate the tensor components, see Section \ref{sec:2}).

We adopt the following approach. It has been said that 
the correlation functions solve the reduced qKZ equation. 
We start with this equation 
and find some class of solutions. 
An arbitrary solution from this class will be 
denoted by $h_n(\la _1,\cdots ,\la _n)$.
The building block of the formula for $h_n$ 
is a family of operators 
\be
\widehat{\Omega}^{(i,j)}_n(\la _1, \cdots ,\la _n)
\in \text{End}
(V^{\otimes 2n})\quad 1\le i<j\le n, 
\en
which are mutually commutative for fixed $\la_1,\cdots,\la_n$ and 
satisfy the equation:
\begin{align}
\widehat{\Omega}^{(i,j)}_n(\la _1, \cdots ,\la _n)
\widehat{\Omega}^{(k,l)}_n(\la _1, \cdots ,\la _n)
=0\quad \text{if}\ \{i,j\}\cap\{k,l\}\ne\emptyset .
\label{nilp}
\end{align}

The solution $h_n\in V^{\otimes 2n}$ has the following form: 
\bea
&&h_n(\la_1,\cdots,\la_n)=2^{-n}
e^{\widehat{\Omega}_n(\la _1,\cdots ,\la _n)}
\bs_{n}, 
\label{main}
\ena
where
$$
\widehat{\Omega}_n(\la _1,\cdots ,\la _n)
=\sum\limits _{i<j}\widehat{\Omega}_n^{(i,j)}
(\la _1,\cdots ,\la _n). 
$$
The expansion for the exponential in \eqref{main}
terminates due to (\ref{nilp}).

The operators $\widehat{\Omega}_n^{(i,j)}$ as functions 
of the parameters $\la _1, \cdots ,\la _n$ are of very special form:
\begin{align}
\widehat{\Omega}^{(i,j)}_n(\la _1, \cdots ,\la _n)=
\ot(\la_i-\la_j)\widetilde{W}^{(i,j)}_n(\z_1,\cdots,\z_n)+
\omega(\la_i-\la_j)W^{(i,j)}_n(\z_1,\cdots,\z_n), 
\end{align}
%
where $\widetilde{W}_n^{(i,j)},W_n^{(i,j)}$ depend rationally on 
$\z_k=e^{\pi i\nu\la_k}$ ($k=1, \cdots ,n$),   
and $\ot(\la),\omega(\la)$ are scalar functions
satisfying a system of difference equations.
Since the series terminates, the
functions $\tilde{\omega}$, $\omega$ 
in the formula for $h_n$ appear 
with total degree not higher than $\left[\frac n 2\right]$.

Finally, the rational functions  $\widetilde{W}^{(i,j)}_n,W^{(i,j)}_n$ are
defined through the trace of a certain monodromy matrix 
(see \eqref{eq:X12}, \eqref{eq:Xij},
\eqref{eq:complicated}, \eqref{eq:X-F}, \eqref{nonsense}).
Here by `trace' we mean the unique linear map 
\be
\Tr_{\la,\z}:U_q(\slt)\otimes \C[\z,\z^{-1}]\to
\la\C[\z,\z^{-1}]\oplus\C[\z,\z^{-1}],
\en
which reduces to the usual trace on the $(k+1)$-dimensional
irreducible representation
when $\la=k+1\in\Z_{> 0}$ and $\z=q^{k+1}$.    
The structure of $\Tr_{\la,\z}$ conforms to the
appearance of {\it two} transcendental functions, 
which is the main new feature in the XXZ case. 
In the body of the text, we use only the difference equation
\eqref{eq:n=2rqKZ} and the parity condition \eqref{eq:n=2rqKZ1} for 
$\ot(\la),\omega(\la)$ and prove that 
the above formula for $h_n$ satisfies the 
reduced qKZ equation and other relations. 

Then we return to correlation functions. 
It has been mentioned that
we describe a class of solutions to the reduced qKZ equation.
Indeed, solutions to difference equations for $\ot(\la),\omega(\la)$ 
are not unique. 
Different  choices of $\ot(\la),\omega(\la)$ 
correspond to different solutions $h_n$. 
We have to choose the one which describes the correlation functions.
The explicit form of $\ot(\la),\omega(\la)$ are determined 
from the case $n=2$ in each regime (see \eqref{eq:om}--\eqref{eq:rt}).  
Actually, for this particular 
solution which we call $h^{\it Ansatz}_n$ the functions 
$\tilde{\omega}(\la)$, $\omega(\la)$  are `minimal': 
they are bounded at infinity having  
the minimal possible number of poles in a certain strip.

It remains to show that our solution describes the correlation functions:
$$
h^{\it Ansatz}_n=h^{\it Corr}_n\,.
$$
In the present paper we prove that this is indeed the case 
in the massive regime. 
Although we were not able to prove this fact in the massless regime,  
we have no doubt that we found the correct solution for the following
reasons. 
First, $h^{\it Ansatz}_n$ 
coincides with $h^{\it Corr}_n$ which was calculated
by other means (starting with original multi-dimensional integral 
formulae \cite{JM}) for $n=2,3$.
Second, $h_n^{\it Ansatz}$ reproduces the correct result in the 
XXX limit. 

Concerning the XXX model it should be said that a formula
similar to (\ref{main}) can be obtained by taking 
the limit from the XXZ case.
The operator  $\widehat{\Omega}^{(i,j)}_n$  simplifies as follows:
\begin{align}
\widehat{\Omega}^{(i,j)\ XXX}_n(\la _1, \cdots ,\la _n)=
\omega ^{XXX}(\la_i-\la_j)W^{(i,j)\ XXX}_n(\la _1,\cdots,\la_n)\nn
\end{align}
where $W^{(i,j)\ XXX}_n(\la _1,\cdots,\la_n)$ is a rational function of 
$\la_1,\cdots, \la _n$ (see (\ref{WXXX})).
Thus the
formula (\ref{main}) gives a new representation even for the XXX 
model. 

We come to the final formula \eqref{main}
in several steps. 
First, we present it using operators 
${}_n\Omega^{(i,j)}_{n-2}(\la_1,\cdots,\la_n)$
which belong to $\Hom(V^{\otimes 2(n-2)},V^{\otimes 2n})$. 
This presentation is quite useful for calculations with small $n$.
We prove that $h_n$ in this form satisfies the reduced qKZ equation
as well as some other relations.
We then show that the result can be rewritten in the 
compact form \eqref{main}.

The plan of the text follows the logic explained above. 
In Section \ref{sec:2},
we formulate the reduced qKZ equation and related properties 
satisfied by the correlation functions. 
In Section \ref{sec:construction_of_h}, 
we introduce operators ${}_nX^{(i,j)}_{n-2}$ 
which belong to $\Hom(V^{\otimes2(n-2)},V^{\otimes2n})\otimes
\left((\la_i-\la_j)\C[\zeta_1^{\pm1},\ldots,\zeta_n^{\pm1}]
\oplus\C[\zeta_1^{\pm1},\ldots,\zeta_n^{\pm1}]\right)$
and use them to construct  
${}_n\Omega^{(i,j)}_{n-2}$.
In Sections \ref{sec:properties-of-X} and 
\ref{sec:preliminary}, 
we describe the main properties of 
${}_nX^{(i,j)}_{n-2}$, from which the reduced qKZ equation
and other properties for $h_n$ can be deduced. 
Sections 
\ref{sec:exchange}--\ref{sec:Recurrence}
are devoted to the proof of the statements
given in Section \ref{sec:preliminary}. 
In Section \ref{sec:another} 
we prove the compact formula (\ref{main})
for the $h_n$.
In Section 
\ref{sec:Correlation}, we discuss the connection 
between the formula given in Section 
\ref{sec:construction_of_h} and the correlation functions.
Theorem \ref{thm:main}, Theorem \ref{thm:e^O} and 
Theorem \ref{thm:massive}
are the main results of this paper.

\section{Reduced qKZ equation}\label{sec:2}

Throughout the paper we fix a complex number $\nu\not \in\Q$.  

Let $V=\C^2$ be a two-dimensional vector space with basis $v_+,v_-$.
We use the standard trigonometric $R$ matrix 
\bea
&&R(\la)=\rho(\la)\frac{r(\la)}{[\la+1]}
\quad \in \End(V\otimes V), 
\label{eq:R}
\\
&&
r(\la)=
\begin{pmatrix}
[\la+1] &      &      &        \\
        &[\la] &   1  &        \\
        &    1 & [\la]&        \\
        &      &      &[\la+1] \\
\end{pmatrix},   
\label{eq:r}
\ena
where 
\be
&&[\la]:=\frac{\sin\pi\nu\la}{\sin\pi\nu}\,.
\en
In \eqref{eq:r}, the entries are arranged according to the order 
$v_+\otimes v_+,v_+\otimes v_-,v_-\otimes v_+,v_-\otimes v_-$. 
The factor $\rho(\la)$ is a meromorphic function subject to the relations 
\be
\rho(\la)\rho(-\la)=1,
\quad \rho(\la)\rho(\la-1)=\frac{[\la]}{[1-\la]}\,.
\en
Its explicit form is irrelevant here (see \eqref{eq:rho}). 
We set also 
\be
\Rc(\la)=P R(\la), 
\en
where $P\in \End(V\otimes V)$ 
stands for the permutation $P(u\otimes v)=v\otimes u$. 

For an element $x\in \End(V)$, 
we set $x_j=\id^{\otimes (j-1)}
\otimes x\otimes \id^{\otimes (N-j)}\in \End(V^{\otimes N})$. 
Similarly, if $y=\sum_\nu y^{(1)}_\nu\otimes\cdots\otimes 
y^{(k)}_{\nu}\in\End(V^{\otimes k})$, 
we let $y_{i_1,\cdots,i_k}=\sum_{\nu}
{\bigl(y^{(1)}_\nu\bigr)}_{i_1}\cdots
{\bigl(y^{(k)}_\nu\bigr)}_{i_k}
\in\End(V^{\otimes N})$. 
An element $\sigma$ of the symmetric group $S_N$ acts on $V^{\otimes N}$ 
as $(v_1\otimes\cdots\otimes v_N)^\sigma
=v_{\sigma^{-1}(1)}\otimes\cdots\otimes v_{\sigma^{-1}(N)}$. 
If $u\in V^{\otimes k}$, $v\in V^{\otimes l}$ and 
$$
\sigma=\begin{pmatrix} 1&\cdots&k&k+1&\cdots k+l\\
i_1&\cdots&i_k&j_1&\cdots j_l\\
\end{pmatrix},
$$
we write $(u\otimes v)^\sigma$ as $u_{i_1,\cdots,i_k}v_{j_1,\cdots,j_l}$.  
Often we fix $n$ and consider the tensor power $V^{\otimes 2n}$. 
When there is no fear of confusion we will write $\bar j=2n-j+1$. 

Consider the following system of equations 
for a sequence $\{h_n\}_{n=0}^\infty$ 
of unknown functions 
$h_n(\la_1,\cdots,\la_n)\in V^{\otimes 2n}$ with $h_0=1$:
\bea
&&
h_n(\cdots,\la_{j+1},\la_j,\cdots)
\label{eq:Rsymm}
\\
&&{}=
\Rc_{j,j+1}(\la_{j,j+1})
\Rc_{\overline{j+1},\bar{j}}(\la_{j+1,j})
h_n(\cdots,\la_j,\la_{j+1},\cdots)
\quad (1\le j\le n-1),\nn
\\
&&
h_n(\la_1-1,\la_2,\cdots,\la_n)
=A^{(1)}_n(\la_1,\cdots,\la_n)
h_n(\la_1,\la_2,\cdots,\la_n),
\label{eq:rqKZ}
\\
&&
\cP^-_{1,\bar{1}}\cdot 
h_n(\la_{1},\cdots,\la_n)
=\frac{1}{2}s_{1,\bar{1}} \,
h_{n-1}(\la_2,\cdots,\la_n)_{2, \cdots , n, \bar{n}, \cdots , \bar{2}}.
\label{eq:n_to_n-1}
\ena
Here $\la_{i,j}=\la_i-\la_j$, $\cP^{-}:=\frac{1}{2}(1- P)$,  
\bea
s:=v_+\otimes v_--v_-\otimes v_+\,\in V\otimes V, 
\label{eq:singl}
\ena
and 
\bea
&&
A^{(1)}_n(\la_1,\cdots,\la_n)
\label{FUNA}
\\
&&=(-1)^n
R_{\bar1,\bar{2}}(\lambda_{1,2}-1)\cdots R_{\bar1,\bar{n}}(\lambda_{1,n}-1)
P_{1,\bar1}
R_{1,n}(\lambda_{1,n})\cdots R_{1,2}(\lambda_{1,2})
\nn
\\
&&=-\prod_{p=2}^n\frac{1}{[\la_{1,p}-1][\la_{1,p}+1]}
\nn\\
&&\quad\times
r_{\bar1,\bar{2}}(\lambda_{1,2}-1)\cdots r_{\bar1,\bar{n}}(\lambda_{1,n}-1)
P_{1,\bar1}
r_{1,n}(\lambda_{1,n})\cdots r_{1,2}(\lambda_{1,2})
\, .
\nn
\ena
Define an operator ${}_{n-1}\Pi_{n} \in \Hom(V^{\otimes 2n}, V^{\otimes 2(n-1)})$ by 
\begin{eqnarray} 
\mathcal{P}_{1, \bar{1}}^{-}u=
s_{1, \bar{1}}({}_{n-1}\Pi_{n}u)_{2, \ldots , n, \bar{n}, \ldots , \bar{2}}. 
\label{eq:def-Pi} 
\end{eqnarray} 
Then \eqref{eq:n_to_n-1} can be written as 
\begin{eqnarray*} 
{}_{n-1}\Pi_{n}\cdot h_{n}(\lambda_{1}, \cdots , \lambda_{n})=
\frac{1}{2}h_{n-1}(\lambda_{2}, \cdots , \lambda_{n}). 
\end{eqnarray*}

These equations arise as ones satisfied by 
the correlation functions of the XXZ model \cite{IIJMNT}.
We review this connection in Section \ref{sec:Correlation}.
Eqs. \eqref{eq:Rsymm}, \eqref{eq:rqKZ} imply 
a reduced form of the usual quantum Knizhnik-Zamolodchikov 
equation of `level $-4$': 
\bea
&&
h_n(\cdots, \la_j-1,\cdots)
=A^{(j)}_n(\la_1,\cdots,\la_n)h_n(\cdots,\la_j,\cdots),
\label{eq:rqKZj}
\ena
where
\bea
\quad&& A^{(j)}_n(\la_1,\cdots,\la_n)
\label{eq:Aj}\\
&&
=(-1)^n
R_{j,j-1}(\la_{j,j-1}-1)\cdots R_{j,1}(\la_{j,1}-1)
R_{\bar j,\overline{j+1}}(\la_{j,j+1}-1)\cdots R_{\bar j,\bar n}(\la_{j,n}-1)
\nn\\
&&\quad\times
P_{j,\bar j}
R_{j,n}(\la_{j,n})\cdots R_{j,j+1}(\la_{j,j+1})
R_{\bar j,\bar1}(\la_{j,1})\cdots 
R_{\bar j,\overline{j-1}}(\la_{j,j-1})\,. 
\nn
\ena
We call \eqref{eq:rqKZj}
the reduced qKZ equation. 
The constraint \eqref{eq:n_to_n-1} 
is compatible in the sense that, 
if $h_n(\la_1,\cdots,\la_n)$ satisfies the 
\eqref{eq:rqKZj}, 
then so does $\cP^-_{1,\bar 1}h_n(\la_1,\cdots,\la_n)$ 
with respect to $(\la_2,\cdots,\la_{n})$
and with $n$ replaced by $n-1$. 

Assigning a weight $\wt (v_{\pm} )=\pm 1$,  
we have the decomposition into weight subspaces 
$V^{\otimes 2n}=\oplus_{l\in\Z}
(V^{\otimes 2n})_l$.  
The reduced qKZ equation is satisfied by 
each weight component of $h_n$. 
In this paper
we consider only the component of zero weight,  
$h_n\in (V^{\otimes 2n})_0$, 
since this is the case relevant to correlation functions. 

Viewed as a system of difference equations,  
our reduced qKZ equation is highly reducible. 
In the first non-trivial case $n=2$, 
there exist functions 
$\tilde{f}(\la_1,\la_2),f(\la_1,\la_2)\in V^{\otimes 4}$
depending rationally on $\zeta_j=e^{\pi i \nu \la_j}$
such that the transformation
\bea
h_2(\la_1,\la_2)=
\ot(\la_{1,2})\tilde{f}(\la_1,\la_2)
+\omega(\la_{1,2})f(\la_1,\la_2)
+1\cdot \frac{1}{4}s_{1,\bar 1}s_{2,\bar 2}
\label{eq:h2}
\ena
brings \eqref{eq:Rsymm}--\eqref{eq:rqKZ} into 
the triangular form, 
\bea
\begin{pmatrix}
\ot(\la-1)\\
\omega(\la-1)\\
1\\
\end{pmatrix}
+
\begin{pmatrix}
1 & 1 & p(\la)+\pt(\la) \\
 0  &1 & p(\la)    \\
 0  & 0  &  -1       \\
\end{pmatrix}
\begin{pmatrix}
\ot(\la)\\
\omega(\la)\\
1\\
\end{pmatrix}
=0,
\label{eq:n=2rqKZ}
\ena
where 
\bea
&&\pt(\la)
:=\frac{1}{2}\frac{1}{[\la-1][\la-2]}
-\frac{1}{4}\frac{[2]}{[\la-1][\la+1]},
\label{eq:pt}\\
&&p(\la):=
\frac{3}{4}\frac{1}{[\la][\la-1]}
-\frac{1}{4}\frac{[3]}{[\la-2][\la+1]},
\label{eq:p}
\ena
together with the parity condition:
\begin{align}
\ot (\la )=-\ot (-\la),\quad \omega (\la) =\omega (-\la).\label{eq:n=2rqKZ1}
\end{align}
Solutions of \eqref{eq:n=2rqKZ} relevant to 
the XXZ model will be 
given in Section \ref{sec:Correlation}, 
\eqref{eq:om}--\eqref{eq:rt}. 

Remarkably, 
the formula \eqref{eq:h2} generalizes for any $n$ as follows:
\bea
h_n(\la_1,\cdots,\la_n)
=\sum_{I,J}\ot_I\omega_J \cdot f_{I,J},
\label{eq:hnrough}
\ena
where $\ot_I=\prod_{(i,j)\in I}\ot(\la_{i,j})$, 
$\omega_J=\prod_{(i,j)\in J}\omega(\la_{i,j})$, 
$f_{I,J}$ are rational functions in $\zeta_1,\cdots,\zeta_n$
and the sum ranges over
$I=(i_1,\cdots,i_m),J=(j_1,\cdots,j_m)\in\{1,\ldots,n\}^m$, 
such that $i_1,\ldots,i_m,j_1,\ldots,j_m$ are distinct,
$i_p<j_p$ $(1\leq p\leq m)$ and $i_1<\cdots<i_m$.
A precise formulation is stated in Theorem \ref{thm:main}. 
In this paper, we prove the existence of solutions 
to \eqref{eq:Rsymm}--\eqref{eq:n_to_n-1}
of the form \eqref{eq:hnrough},  
by constructing the rational functions $f_{I,J}$ explicitly.

\section{Construction of $h_n$}\label{sec:construction_of_h}

In this section we introduce 
the solution $h_n$ and state the result. 

In the following construction, a key role will 
be played by a family of linear maps
\be
{}_{n}X^{(i,j)}_{n-2}(\la_1,\cdots,\la_n)
\quad
\in 
\Hom(V^{\otimes 2(n-2)},V^{\otimes 2n})\,
\quad (1\le i<j\le n) 
\en
depending on the parameters $\la_1,\cdots,\la_n$. 
They can be viewed as certain transfer matrices 
whose auxiliary spaces have `fractional dimension'.
We begin by defining these objects. 

Let $q=e^{\pi i\nu}$. 
Let $E,F,q^{\pm H/2}$ be the standard generators of $U_q(\slt)$. 
Define the $L$ operator by the formula
\bea
L(\la):=
\begin{pmatrix}
\ds{\Bigl[\la+\frac{1+H}{2}\Bigr]}&\tilde{F} \\
\tilde{E}          &\ds{\Bigl[\la+\frac{1-H}{2}\Bigr]} \\
\end{pmatrix}
\quad\in U_q(\slt)\otimes\End(V), 
\label{eq:L}
\ena
where $\tilde{F}=Fq^{(H-1)/2}$, $\tilde{E}=q^{-(H-1)/2}E$. 
Denote by $\pi^{(k)}:U_q(\slt)\to \End(V^{(k)})$ 
the $(k+1)$-dimensional irreducible representation. 
There exists a unique $\C[\z,\z^{-1}]$-linear map
\be
\Tr_{\la,\z}:U_q(\slt)\otimes\C[\z,\z^{-1}]
\longrightarrow 
\la\C[\z,\z^{-1}]\oplus\C[\z,\z^{-1}]
\en
such that 
\be
\Tr_{\la,\z}\bigl(A\bigr)\Bigl|_{\la=k+1,\z=q^{k+1}}
=
\tr_{V^{(k)}}\left(\pi^{(k)}(A)\right)
\quad (\forall k\in\Z_{\ge 0})
\en
holds for all $A\in U_q(\slt)$. 
For the properties of $\Tr_{\la,\z}$, 
see Section \ref{sec:preliminary}. 
An element $f(\la,\z)\in\C[\la,\z,\z^{-1}]$ is 
determined by knowing $f(\la,q^\la)$ 
(see Lemma \ref{lem:integer} below).  
For this reason we will mainly consider 
\be
\Tr_\la(A):=\Tr_{\la,q^\la}(A).
\en

Notation being as above, we define 
\bea
&&
{}_{n}X_{n-2}(\la_1,\cdots,\la_n)(u)
\label{eq:X12}\\
&&
:=
\frac1{[\lambda_{1,2}]\prod_{p=3}^n[\lambda_{1,p}][\lambda_{2,p}]}
\Tr_{\lambda_{1,2}}
\Bigr(T^{[1]}_n\bigl(\frac{\lambda_1+\lambda_2}{2}\bigr)\Bigl)
(s_{1,\bar{2}}s_{\bar{1},2}u_{3,\cdots,n,\bar n,\cdots,\bar 3})
\,.
\nn
\ena
Here $u\in V^{\otimes 2(n-2)}$, 
$s$ is given in \eqref{eq:singl}, 
and 
$T^{[i]}_n(\la)\in  U_q(\slt)\otimes\End(V^{\otimes2n})$
stands for the `monodromy matrix'
\be
T^{[i]}_n(\lambda;\la_1,\ldots,\la_n)
&=&
L_{\bar1}(\la-\la_1-1)\cdots 
\widehat{L_{\bar i}(\la-\la_i-1)}
\cdots L_{\bar n}(\la-\la_n-1)
\\
&\times&
L_n(\la-\la_n)\cdots 
\widehat{L_{i}(\la-\la_i)}
\cdots L_1(\la-\la_1). 
\en
In the above, we have abbreviated 
$T^{[i]}_n(\lambda;\la_1,\ldots,\la_n)$ to $T^{[i]}_n(\lambda)$.
In general, we define 
\bea
&&
{}_{n}X_{n-2}^{(i,j)}(\la_1,\cdots,\la_n)
\label{eq:Xij}\\ &&
\quad=\mathbb{R}_n^{(i,j)}(\la_1,\cdots,\la_n)
\circ {}_{n}X_{n-2}(\la_i,\la_j,\la_1,
\cdots,\widehat{\la_i},\cdots,\widehat{\la_j},\cdots,\la_n), \nn
\ena
where
\begin{eqnarray}
&&\mathbb{R}_n^{(i,j)}(\la_1,\cdots,\la_n)
\label{eq:complicated}
\\
&&:=
\Rc_{i,i-1}(\lambda_{i,i-1})\cdots \Rc_{2,1}(\lambda_{i,1}) 
\nn\\ 
&& {}\times 
\Rc_{j,j-1}(\lambda_{j,j-1})\cdots 
\Rc_{i+2,i+1}(\lambda_{j,i+1})\cdot
\Rc_{i+1,i}(\lambda_{j,i-1})\cdot
\cdots \Rc_{3,2}(\lambda_{j,1})
\nn\\
&&{}\times
\Rc_{\overline{i-1},\bar i}(\lambda_{i-1,i})\cdots
\Rc_{\bar 1\bar 2}(\lambda_{1,i}) 
\nn\\ 
&& {}\times 
\Rc_{\overline{j-1},\bar{j}}(\lambda_{j-1,j})\cdots 
\Rc_{\overline{i+1},\overline{i+2}}(\lambda_{i+1,j})
\cdot 
\Rc_{\overline{i},\overline{i+1}}(\lambda_{i-1,j})
\cdots \Rc_{\bar{2},\bar{3}}(\lambda_{1,j}).
\nn
\end{eqnarray}
The complicated expression of 
$\mathbb{R}_n^{(i,j)}(\la_1,\cdots,\la_n)$
can be understood as a chain of processes of reversing the order of arguments.
(See \eqref{eq:X-Rsym}.)
Using the Yang-Baxter relation
\bea
R_{1,2}(\la_{1,2})L_2(\la-\la_2)L_1(\la-\la_1)
=
L_1(\la-\la_1)L_2(\la-\la_2)R_{1,2}(\la_{1,2}), 
\label{eq:YBE}
\ena
it can be written alternatively as 
\be
&&{}_{n}X_{n-2}^{(i,j)}(\la_1,\cdots,\la_n)(u)
=
R_{i,i-1}(\la_{i,i-1})\cdots R_{i,1}(\la_{i,1})
R_{\overline{i-1},\bar i}(\la_{i-1,i})\cdots 
R_{\bar 1,\bar i}(\la_{1,i})
\\
&&\quad \times
\frac{1}{[\la_{i,j}]\prod_{p(\neq i,j)}[\la_{i,p}][\la_{j,p}]}
\Tr_{\la_{i,j}}\left(T^{[i]}_n
\Bigl(\frac{\la_i+\la_j}{2}\Bigr)\right)
\\
&&\quad \times
R_{j,j-1}(\la_{j,j-1})
\cdots
\widehat{R_{j,i}(\la_{j,i})}
\cdots 
R_{j,1}(\la_{j,1})
R_{\overline{j-1},\bar j}(\la_{j-1,j})
\cdots 
\widehat{R_{\overline{i},\bar j}(\la_{i,j})}
\cdots
R_{\bar1,\overline{j}}(\la_{1,j})
\\
&&\quad\times
\left(s_{i,\bar j}s_{\bar i,j}
u_{1,\cdots,\hat{i},\cdots,\hat{j},\cdots,n,
\bar{n},\cdots,\hat{\bar j},\cdots,\hat{\bar{i}},\cdots,\bar{1}}
\right).
\en

By the definition, ${}_{n}X_{n-2}^{(i,j)}$ has the structure
\bea
{}_{n}X_{n-2}^{(i,j)}(\la_1,\cdots,\la_n)=
-\la_{i,j}\cdot{}_{n}\Ft_{n-2}^{(i,j)}(\z_1,\cdots,\z_n)
+{}_{n}\F_{n-2}^{(i,j)}(\z_1,\cdots,\z_n),
\label{eq:X-F}
\ena
where ${}_{n}\Ft_{n-2}^{(i,j)}(\z_1,\cdots,\z_n)$ and 
${}_{n}\F_{n-2}^{(i,j)}(\z_1,\cdots,\z_n)$
are rational functions in $\z_1,\cdots,\z_n$.
The properties of ${}_{n}X^{(i,j)}_{n-2}$ will be discussed in  
Section \ref{sec:properties-of-X}. 

With each solution $\ot(\la),\omega(\la)$ 
of the difference equation \eqref{eq:n=2rqKZ} and the parity condition \eqref{eq:n=2rqKZ1}, 
we now associate 
a $V^{\otimes 2n}$-valued function $h_n$ as follows. 
Set 
\bea
&&{}_{n}\Omega_{n-2}^{(i,j)}(\la_1,\cdots,\la_n)
\label{eq:O-F}\\
&&\quad:=
\ot(\la_{i,j})\cdot{}_{n}\Ft_{n-2}^{(i,j)}(\z_1,\cdots,\z_n)
+\omega(\la_{i,j})\cdot {}_{n}\F_{n-2}^{(i,j)}(\z_1,\cdots,\z_n).
\nn
\ena
For an ordered set of 
indices $K=\{k_1,\cdots,k_m\}$ ($1\le k_1<\cdots<k_m\le n$), 
we use the abbreviated notation 
\bea
\Omega_{K,(k_i,k_j)}=
{}_{m}\Omega^{(i,j)}_{m-2}(\la_{k_1},\cdots,\la_{k_m}).\label{OMK}
\ena
We set also
\be
\bs_m:=\prod_{p=1}^m s_{p,\bar p}.
\en

Define 
\bea
&&h_n(\la_1,\cdots,\la_n)
\label{eq:def-hn}\\
&&:=
\sum_{m=0}^{[n/2]}\frac{(-1)^m}{2^{n-2m}}\sum
\Omega_{K_1,(i_1,j_1)}\circ
\Omega_{K_2,(i_2,j_2)}\circ
\cdots \circ\Omega_{K_{m},(i_m,j_m)}
\left(\bs_{n-2m}\right).  
\nn
\ena
Here the second sum is taken over all 
sequences $i_1,\cdots,i_m,j_1,\cdots,j_m$ 
of distinct elements of $K_1=\{1,\cdots,n\}$ 
such that $i_1<\cdots<i_m$ and $i_1<j_1,\cdots,i_m<j_m$.  
The $K_l$ ($l=2,\cdots,m$) are subsets of $K_1$ 
obtained by removing $i_{l-1},j_{l-1}$ from $K_{l-1}$. 
For instance, if $n=4$, then
\be
&&h_4(\la_1,\cdots,\la_4)=
\sum^{\mbox{\tiny{3 terms}}}
{}_{4}\Omega^{(1,2)}_2(\la_1,\cdots,\la_4)
\circ {}_{2}\Omega^{(1,2)}_0(\la_3,\la_4)(1)
\\
&&\quad -
\frac{1}{4}
\sum^{\mbox{\tiny{6 terms}}}
{}_{4}\Omega^{(1,2)}_2(\la_1,\cdots,\la_4)(\bs_2)
+\frac{1}{16}\bs_4.
\en
Expanding \eqref{eq:def-hn} in terms of 
$\ot(\la_{i,j})$ and $\omega(\la_{i,j})$, 
we obtain an expression of the type 
\eqref{eq:hnrough}. 

We are now in  a position to state the result. 

\begin{thm}\label{thm:main}
Let 
$\ot(\la),\omega(\la)$ be a solution of the difference equation 
\eqref{eq:n=2rqKZ} and the parity condition \eqref{eq:n=2rqKZ1}. 
Define $h_n$ by \eqref{eq:def-hn}, 
where ${}_{n}\Omega^{(i,j)}_{n-2}$ are given by 
\eqref{eq:X-F} and \eqref{eq:O-F}. 
Then $h_n$ satisfies the reduced qKZ equation
\eqref{eq:Rsymm}, \eqref{eq:rqKZ} as well as 
\eqref{eq:n_to_n-1}.
\end{thm}

\medskip

\noindent{\it{Example.}}\\
In the simplest non-trivial case $n=2$, 
we have the following representation for $h_2$: 
\be
&&
h_2(\la_1,\la_2)=
-\ot(\la_{1,2})\cdot {}_2\Ft_0(\z_1,\z_2)(1)
-\omega(\la_{1,2})\cdot {}_2\F_0(\z_1,\z_2)(1)
+\frac{1}{4}s_{1,\bar1}s_{2,\bar2},
\\
&&
(q-q^{-1})^2\cdot {}_2\Ft_0(\z_1,\z_2)(1)=
\frac{q^2-q^{-2}}{\z-\z^{-1}}
A+
\frac{(\z+\z^{-1})(q-q^{-1})}{\z-\z^{-1}}
B,
\\
&&
(q-q^{-1})^2\cdot {}_2 G_0(\z_1,\z_2)(1)=
(\z+\z^{-1})
A+
(q+q^{-1})
B,
\\
&&A=(+--+)+(-++-),
\\
&&B=(++--)+(--++)+(+-+-)+(-+-+),
\en
Here we have set $\z=\z_1/\z_2$ and 
$(\epsilon_1,\epsilon_2,\overline{\epsilon}_2, 
\overline{\epsilon}_1)
=v_{\epsilon_1}\otimes
v_{\epsilon_2}\otimes
v_{\overline{\epsilon}_2}\otimes
v_{\overline{\epsilon}_1}$.

\vspace{0.2cm}

The following sections are devoted to the proof of the theorem.

\section{Reduction of reduced qKZ equation}
\label{sec:properties-of-X}

The system of equations \eqref{eq:Rsymm}, \eqref{eq:rqKZ}, \eqref{eq:n_to_n-1} 
for $h_n$ is based on the following five properties.


\begin{description}
\item[Exchange relation]
\bea
&&
{}_{n}X_{n-2}^{(i,j)}(\cdots,\la_{k+1},\la_k,\cdots)
=
\Rc_{k,k+1}(\la_{k,k+1})
\Rc_{\overline{k+1},\overline{k}}(\la_{k+1,k})
\label{eq:X-Rsym}
\ena
\bea
&& 
\times
\begin{cases}
{}_{n}X_{n-2}^{(k,k+1)}(\cdots,\la_{k},\la_{k+1},\cdots), 
& (i=k,j=k+1),  \\
{}_{n}X_{n-2}^{(\pi_k(i),\pi_k(j))}(\cdots,\la_{k},\la_{k+1},\cdots), 
& ( \sharp(\{i,j\}\cap\{k,k+1\})=1), 
\\
{}_{n}X_{n-2}^{(i,j)}(\cdots,\la_{k},\la_{k+1},\cdots) & {}
\\ \qquad {}\times 
\Rc_{k',k'+1}(\la_{k,k+1})^{-1}
\Rc_{\overline{k'+1},\overline{k'}}(\la_{k+1,k})^{-1}, 
& (\{i,j\}\cap\{k,k+1\}=\emptyset).
\\
\end{cases}
\nn
\ena
\end{description}
Here $\pi_k$ signifies the transposition $(k,k+1)$, 
and in the last line $k'$ signifies
the position of $\lambda_k$ appearing in 
$(\lambda_{1}, \cdots , \widehat{\lambda_{i}}, \cdots , \widehat{\lambda_{j}}, \cdots , \lambda_{n})$. 
\begin{description}
\item[Difference equations] 
\bea
&&
{}_{n}X_{n-2}^{(i,j)}(\la_1,\cdots,\la_k-1,\cdots,\la_n)
\label{eq:X-diff1}\\
&&
=-A^{(k)}_n(\la_1,\cdots,\la_n)\circ {}_{n}X_{n-2}^{(i,j)}(\la_1,\cdots,\la_n)
\quad  (k=i,j), 
\nn
\\
&&
{}_{n}X_{n-2}^{(i,j)}(\la_1,\cdots,\la_k-1,\cdots,\la_n)
=A^{(k)}_n(\la_1,\cdots,\la_n)\circ {}_{n}X_{n-2}^{(i,j)}(\la_1,\cdots,\la_n)
\label{eq:X-diff2}\\
&&{}\quad \times
A^{(k')}_{n-2}(\la_1,\cdots,\widehat{\la_i},\cdots,\widehat{\la_j},
\cdots,\la_n)^{-1}
\quad (k\neq i,j).
\nn
\ena
\end{description}
Here $k'$ signifies
the position of $\lambda_k$ appearing in 
$(\lambda_{1}, \cdots , \widehat{\lambda_{i}}, \cdots , \widehat{\lambda_{j}}, \cdots , \lambda_{n})$. 
\begin{description}
\item[Commutativity]
For distinct indices $i,j,k,l$, 
\bea
&&
{}_{n}X_{n-2}^{(i,j)}(\la_1,\cdots,\la_n)
\circ {}_{n-2}X_{n-4}^{(k',l')}
(\la_1,\cdots,\widehat{\la_{i}},\cdots,\widehat{\la_{j}},\cdots,\la_n)
\label{eq:X-comm}\\
&&
=
{}_{n}X_{n-2}^{(k,l)}(\la_1,\cdots,\la_n)
\circ {}_{n-2}X_{n-4}^{(i',j')}
(\la_1,\cdots,\widehat{\la_{k}},\cdots,
\widehat{\la_{l}},\cdots,\la_n). 
\nn
\ena
\end{description}
Here $k'$ and $l'$ signify 
the positions of $\lambda_k$, $\lambda_{l}$ appearing in 
$(\lambda_{1}, \cdots , \widehat{\lambda_{i}}, \cdots , \widehat{\lambda_{j}}, \cdots , \lambda_{n})$. 
Likewise $i'$ and $j'$ signify the positions of $\lambda_{i}$, $\lambda_{j}$ appearing in 
$(\lambda_{1}, \cdots , \widehat{\lambda_{k}}, \cdots , \widehat{\lambda_{l}}, \cdots , \lambda_{n})$. 
\begin{description}
\item[Cancellation identity]
\bea
&&
4\sum_{j=2}^n
{}_{n}Y_{n-2}^{(1,j)}(\z_1,\cdots,\z_n)
\bigl(\bs_{n-2}\bigr)
=\left(A^{(1)}_n(\la_1,\cdots,\la_n)^{-1}-1\right)
(\bs_n).
\label{eq:X-last}
\ena
\end{description}
\begin{description}
\item[Recurrence relation]
\bea
&& 
\label{eq:X-red} 
{}_{n-1}\Pi_{n}\circ
{}_{n}
X_{n-2}^{(i,j)}(\la_1,\cdots,\la_n)\\
&&=
\begin{cases}
0 & (1=i<j\le n), \\
{}_{n-1}X_{n-3}^{(i-1,j-1)}(\la_2,\cdots,\la_n) 
\circ {}_{n-3}\Pi_{n-2} & (2\le i<j\le n). \\
\end{cases}
\nonumber
\ena
\end{description}
In \eqref{eq:X-last}, we have introduced
the function 
\bea
&&{}_{n}Y^{(i,j)}_{n-2}(\z_1,\cdots,\z_n)
:=\pt(\la_{i,j})\cdot {}_{n}\Ft_{n-2}^{(i,j)}(\z_1,\cdots,\z_n)
+p(\la_{i,j})\cdot{}_{n} \F_{n-2}^{(i,j)}(\z_1,\cdots,\z_n),
\label{eq:Oc}
\ena
which is rational in $\z_1,\cdots,\z_n$. 

\begin{prop}
The properties \eqref{eq:X-Rsym}--\eqref{eq:X-red}
imply the reduced qKZ equation 
\eqref{eq:Rsymm}, 
\eqref{eq:rqKZ} and \eqref{eq:n_to_n-1} 
for $h_n$.
\end{prop}

\begin{proof}
Eqs.\eqref{eq:X-Rsym}, \eqref{eq:X-diff2} and \eqref{eq:X-comm} 
are equivalent to the corresponding 
set of relations for the rational 
functions ${}_{n}\Ft^{(i,j)}_{n-2}$ and ${}_{n}\F^{(i,j)}_{n-2}$. 
Hence they hold true if we replace ${}_{n}X^{(i,j)}_{n-2}$ by
${}_{n}\Omega^{(i,j)}_{n-2}$. 
The exchange symmetry \eqref{eq:Rsymm} for 
$h_n$ is obvious from this remark.  

Let us consider \eqref{eq:rqKZ}. 
It follows from \eqref{eq:n=2rqKZ} and \eqref{eq:X-diff1} that 
the definition \eqref{eq:Oc} of ${}_{n}Y^{(1,j)}_{n-2}$ can be rewritten as
\bea
&&{}_{n}\Omega_{n-2}^{(1,j)}(\la_1-1,\cdots,\la_n)
\label{eq:Y-diff}\\
&&\quad=
A^{(1)}_n(\la_1,\cdots,\la_n)
\left({}_{n}\Omega_{n-2}^{(1,j)}(\la_1,\cdots,\la_n)
+{}_{n}Y_{n-2}^{(1,j)}(\z_1,\cdots,\z_n)\right). 
\nn
\ena
Set 
\be
&&\phi_n(\la_1,\cdots,\la_n):=
-
4\sum_{j=2}^n{}_{n}\Omega^{(1,j)}_{n-2}(\la_1,\cdots,\la_n)
\bigl(\bs_{n-2}\bigr)
+
\bs_n.
\en
Then \eqref{eq:Y-diff} and \eqref{eq:X-last} imply that 
\bea
\phi_n(\la_1-1,\cdots,\la_n)=
A^{(1)}_n(\la_1,\cdots,\la_n)\phi_n(\la_1,\cdots,\la_n).
\label{eq:Z}
\ena
In the definition \eqref{eq:def-hn} of $h_n$, 
the sum is taken over sequences $i_1,\cdots,i_m,j_1,\cdots,j_m$. 
The commutativity \eqref{eq:X-comm} implies that
\bea
\Omega_{K,(1,j)}\circ\Omega_{K\backslash\{1,j\},(k,l)}
=
\Omega_{K,(k,l)}\circ\Omega_{K\backslash\{k,l\},(1,j)}.
\label{comm}
\ena 
Repeating this procedure, 
one can bring the index `$1$' to the last position.  
Collecting terms containing `$1$', we obtain 
\bea
{}\quad 
h_n=
\sum_{m=0}^{[n/2]}\frac{(-1)^m}{2^{n-2m}}\sum
\Omega_{K_1,(i_1,j_1)}\circ
\Omega_{K_2,(i_2,j_2)}\circ
\cdots \circ\Omega_{K_{m},(i_m,j_m)}
\left(\phi_{n-2m}\right),  
\label{eq:hn-1}
\ena
where the sum is taken by the same rule as in \eqref{eq:def-hn}
except that $i_l,j_l\neq 1$ ($l=1,\cdots,m$). 
The reduced qKZ equation \eqref{eq:rqKZ} for $h_n$ 
is an immediate consequence of \eqref{eq:hn-1}, \eqref{eq:X-diff2},
and \eqref{eq:Z}. 

Finally \eqref{eq:n_to_n-1} is a simple consequence of
the relation \eqref{eq:X-red}.
\end{proof}
\medskip

\section{Properties of ${}_{n}X^{(i,j)}_{n-2}$}
\label{sec:preliminary}

Let us study some simple properties of ${}_{n}X^{(i,j)}_{n-2}$. 

The $\C[\z,\z^{-1}]$-linear map
\be
\Tr_{\la,\z}:U_q(\slt)\otimes \C[\z,\z^{-1}]
\longrightarrow
\la\C[\z,\z^{-1}]\oplus \C[\z,\z^{-1}]
\en
is uniquely characterized by the following properties.
\bea
&&\Tr_{\la,\z}(AB)=\Tr_{\la,\z}(BA),
\label{eq:tr1}
\\
&&\Tr_{\la,\z}(A)=0\qquad \mbox{if $A$ has non-zero weight},
\label{eq:tr2}\\
&&\Tr_{\la,\z}(q^{mH})=
\begin{cases}
\la & (m=0), \\
\frac{\z^m-\z^{-m}}{q^m-q^{-m}} & (m\neq 0),\\
\end{cases}
\label{eq:tr3}\\
&&\Tr_{\la,\z}(C A)
=\frac{\z+\z^{-1}}{(q-q^{-1})^2}\Tr_{\la,\z}(A),
\label{eq:tr4}
\ena
where $A,B\in U_q(\slt)$, and 
\be
C=\frac{q^{-1+H}+q^{1-H}}{(q-q^{-1})^2}+EF
\en
is a central element. 
It follows that for any $A\in U_q(\slt)$ we have 
\be
\Tr_{k+1,q^{k+1}}(A)
=\tr_{V^{(k)}}\bigl(\pi^{(k)}(A)\bigr)
\quad (k\in\Z_{\ge 0}), 
\en
and
\bea
\Tr_{-\la,\z^{-1}}(A)=-\Tr_{\la,\z}(A).
\label{eq:odd}
\ena

For $k\in\Z_{\ge 0}$, we set
\be
r^{(k,1)}(\la):=
\bigl(\pi^{(k)}\otimes\id\bigr)\bigl(L(\la)\bigr). 
\en
We have $r^{(1,1)}(\la)=r(\la)$. 
We will often use crossing symmetry
\bea
L_{a}(\la)s_{a,b}=-L_b(-\la-1)s_{a,b}.
\label{eq:crossing}
\ena
Since
\be
L_1\Bigl(\frac{\mu}{2}\Bigr)L_2\Bigl(\frac{\mu}{2}-1\Bigr)s_{1,2}
&=&\left(\frac{q^\mu+q^{-\mu}}{(q-q^{-1})^2}-C\right)
s_{1,2}, 
\en
we have for any $A\in U_q(\slt)$ 
\bea
&&\Tr_\la
\left(A
L_1\Bigl(\frac{\mu}{2}\Bigr)L_2\Bigl(\frac{\mu}{2}-1\Bigr)
\right)s_{1,2}
=
\Bigl[\frac{\mu-\la}{2}\Bigr]\Bigl[\frac{\mu+\la}{2}\Bigr]
\Tr_\la\left(A\right)s_{1,2}.
\label{eq:q-det}
\ena
Note that \eqref{eq:q-det} implies that 
\begin{eqnarray}
{\rm Tr}_{\lambda}\left(
A \, \mathcal{P}^{-}_{1,2}L_{2}(\frac{\mu}{2}-1) L_{1}(\frac{\mu}{2})
\right)&=&
{\rm Tr}_{\lambda}\left(
A \, L_{1}(\frac{\mu}{2})L_{2}(\frac{\mu}{2}-1)\mathcal{P}^{-}_{1,2} 
\right) 
\label{eq:q-det-modified}
\\ 
&=& 
\Bigl[\frac{\mu-\la}{2}\Bigr]\Bigl[\frac{\mu+\la}{2}\Bigr]
\Tr_\la\left(A\right)
\mathcal{P}^{-}_{1,2}. 
\nonumber
\end{eqnarray}

\begin{lem}\label{lem:power_of_trace}
For each $\epsilon_p,\epsilon_p', 
\bar{\epsilon}_p,\bar{\epsilon}_p'
\in \{1,-1\}$ $(2\le p\le n)$,  
set $l=\sum_{p=2}^n(\epsilon_p+\bar{\epsilon}_p)$, 
$l'=\sum_{p=2}^n(\epsilon'_p+\bar{\epsilon}'_p)$. 
Then as a function of $\la_1$ we have 
\bea
&&
\Tr_{\la_{1,2}}\Bigl(
L\bigl(\frac{\la_{1,2}}{2}-1\bigr)_{\bar{\epsilon}_2,\bar{\epsilon}'_2}
L\bigl(\frac{\la_1+\la_2}{2}-\la_3-1\bigr)
_{\bar{\epsilon}_3,\bar{\epsilon}'_3}
\cdots
L\bigl(\frac{\la_1+\la_2}{2}-\la_n-1\bigr)
_{\bar{\epsilon}_n,\bar{\epsilon}'_n}
\label{eq:LLL}
\\
&&\quad \times
L\bigl(\frac{\la_1+\la_2}{2}-\la_n\bigr)
_{\epsilon_n,\epsilon'_n}
\cdots
L\bigl(\frac{\la_1+\la_2}{2}-\la_3\bigr)
_{\epsilon_3,\epsilon'_3}
L\bigl(\frac{\la_{1,2}}{2}\bigr)
_{\epsilon_2,\epsilon'_2}
\Bigr)
\nn\\
&&=\begin{cases}
\zeta_1^{-(n-1+|l|/2)}
(\la_{1,2}\tilde{g}(\zeta_1^2)+g(\zeta_1^2))
& (l=l') \\
0& (l\neq l'),\\
\end{cases}
\nn
\ena
where $\tilde{g}(z),g(z)$ 
are polynomials of degree at most $n-1+|l|/2$. 
\end{lem}
\begin{proof}
Let us write $\bar{\epsilon}_2,\cdots,
\bar{\epsilon}_n,\epsilon_n,\cdots, \epsilon_2$ as 
$\varepsilon_1,\cdots,\varepsilon_{2n-2}$.
The expression inside the trace is a product of factors
of the form $\tilde{E}$, $\tilde{F}$, 
$\Bigl[\frac{\la_{1,2}+\varepsilon_pH}{2}+c_p\Bigr]$,
where $c_p$ is a $\Z$-linear combination of 
$1/2,\la_2,\cdots,\la_n$. 
The $\Tr$ has a non-zero value only when 
$\tilde{E}$ and $\tilde{F}$ occur
the same number of times. 
Hence it is $0$ if $l\neq l'$. 

Consider the case $l=l'$.  
By the property \eqref{eq:tr4} we have  
\be
&&
EF\equiv
\Bigl[\frac{\la+1-H}{2}\Bigr]\Bigl[\frac{\la-1+H}{2}\Bigr],
\\
&&
FE\equiv
\Bigl[\frac{\la+1+H}{2}\Bigr]\Bigl[\frac{\la-1-H}{2}\Bigr],
\en
where $X\equiv X'$ means that $\Tr_\la(AX)=\Tr_\la(AX')$ 
for all $A\in U_q(\slt)$. 
Using them one can reduce \eqref{eq:LLL} 
to a linear combination of terms of the form 
\be
&&\Tr_{\la_{1,2}}
\left(q^{(n-1)\la_{1,2}+(l/2)H-\sum_{p\in I}(\la_{1,2}+\varepsilon_p H)}
\right)
\\
&&=
q^{(n-1)\la_{1,2}-\sharp I\cdot \la_{1,2}}
\times 
\begin{cases}
\frac{[(l/2-\sum_{p\in I}\varepsilon_p)\la_{1,2}]}
{[l/2-\sum_{p\in I}\varepsilon_p]}
& (\sum_{p\in I}\varepsilon_p\neq l/2),\\
\la_{1,2} & (\mbox{otherwise}), \\
\end{cases}
\en
where $I$ is a subset of $\{1,\cdots,2n-2\}$. 
The last expression is a linear combination of 
$\z_1^{s}$ or $\la_{1,2}\z_1^s$, with the possible power
(note that $l$ is even) 
\be
s=n-1\pm\frac{l}{2}-\sum_{p\in I}(1\pm \varepsilon_p).
\en
It is easy to see that $s\equiv n-1+|l|/2\bmod 2$ and 
$|s|\le n-1+|l|/2$. 
\end{proof}

\begin{lem}\label{lem:parity}
The functions ${}_{n}\Ft_{n-2}^{(1,2)}$ has the parity property  
\bea
&&
{}_{n}\Ft_{n-2}^{(1,2)}(\z_1,\cdots,-\z_j,\cdots,\z_n)
\label{eq:F-Parity}\\
&&=\sigma^z_j\sigma^z_{\bar j}\cdot
{}_{n}\Ft_{n-2}^{(1,2)}(\z_1,\cdots,\z_j,\cdots,\z_n)
\cdot
\begin{cases}
(-1)^{n-1}\prod_{p=1}^{n-2}(\sigma^z_p\sigma^z_{\bar p})
& (j=1,2),\\
\sigma^z_{j-2}\sigma^z_{\overline{j-2}}&(3\le j\le n).
\end{cases}\nn
\ena
In particular, we have
\bea
{}\quad 
&&
{}_{n}\Ft_{n-2}^{(1,2)}(\z_1,\cdots,-\z_j,\cdots,\z_n)(\bs_{n-2})
=-\sigma^z_j\sigma^z_{\bar j}\cdot 
{}_{n}\Ft_{n-2}^{(1,2)}(\z_1,\cdots,\z_j,\cdots,\z_n)(\bs_{n-2})
\label{eq:F-Parity2}\ena

These relations hold also for ${}_{n}\F^{(1,2)}_{n-2}$ in place of ${}_{n}\Ft^{(1,2)}_{n-2}$. 
\end{lem}
\begin{proof}

The case $j\ge 3$ is a consequence of the relation
\bea
L(\la+1/\nu)=-\sigma^zL(\la)\sigma^z.
\label{eq:L-parity}
\ena
For $j=1$, \eqref{eq:F-Parity}
follows from Lemma \ref{lem:power_of_trace} and 
\be
\prod_{p=1}^n\sigma^z_p\sigma^z_{\bar p}
\cdot{}_{n}X^{(1,2)}_{n-2}(\la_1,\cdots,\la_n)
={}_{n}X^{(1,2)}_{n-2}(\la_1,\cdots,\la_n)
\cdot\prod_{p=1}^{n-2}\sigma^z_p\sigma^z_{\bar p}.
\en 

Finally the case 
with $j=2$ follows from 
the translation invariance 
\begin{eqnarray*}
{}_{n}X_{n-2}^{(i, j)}(\lambda_{1}+\mu, \ldots , \lambda_{n}+\mu)=
{}_{n}X_{n-2}^{(i, j)}(\lambda_{1}, \ldots , \lambda_{n}) 
\end{eqnarray*}
and the rest of \eqref{eq:F-Parity} 
by simultaneously shifting $\la_j$ by $1/\nu$.
\end{proof}

Define an operator acting on $V^{\otimes2n}$ by 
\begin{eqnarray}
t^{[1,\cdots,k]}_{n,a}(\la; \lambda_{k+1}, \ldots , \lambda_{n})
&:=&R_{a,\overline{k+1}}(\la-\la_{k+1}-1)
\cdots R_{a,\bar{n}}(\la-\la_{n}-1)
\label{eq:t_a}\\
&\quad&\times
R_{a,n}(\la-\la_{n})\cdots
R_{a,k+1}(\la-\la_{k+1})
\nn 
\end{eqnarray}
for $a\in\{1,\ldots,k,\bar k,\ldots,\bar1\}$.
The following specializations are sometimes useful. 

\begin{lem}\label{lem:special}
We have
\bea
&&{}_{n}X_{n-2}^{(i,j)}(\la_1,\cdots,\la_n)\Bigl|_{\la_{i,j}=\pm 1}=0,
\label{eq:Xla=1}\\
&&
{}_{n}X_{n-2}^{(1,2)}(\la_1,\cdots,\la_n)\Bigl|_{\la_{1,2}=-2}(u)
=(-1)^{n-1}\frac{1}{[2]}r_{\bar1,\bar2}(-2)
\label{eq:Xla=2}
\\
&&\quad\times 
t^{[1,2]}_{n,\bar 1}
(\la_2-1;\la_3,\cdots,\la_n)
\bigl(s_{1,\bar2}s_{\bar 1,2}
u_{3,\cdots,n,\bar n,\cdots,\bar 3}\bigr),
\nn
\ena
\end{lem}
\begin{proof}
By the definition we have 
$\Tr_1(A)=\varepsilon(A)$, where
$\varepsilon:U_q(\slt)\to\C$ stands for the counit.  
Since 
\be
\varepsilon\left(T^{[i]}_n\bigl(\frac{\la_i+\la_j}{2}\bigr)\right)
\en
is divisible by $[\frac{\la_{i,j}+1}2][\frac{\la_{i,j}-1}2]$, we obtain
\eqref{eq:Xla=1}. 
Similarly \eqref{eq:Xla=2} follows by noting 
$\tr_{V_a}(A_a r_{a,b}(0))=A_b$. 
\end{proof}

\begin{lem}\label{lem:pole}
The coefficients of the linear map
\bea
[\la_{i,j}]\prod_{p(\neq i,j)}^n[\la_{i,p}][\la_{j,p}]
\cdot
{}_{n}X_{n-2}^{(i,j)}(\la_1,\cdots,\la_n)
\label{eq:X-pole}
\ena
belong to
$\la_{i,j}\C[\zeta_1^{\pm1},\ldots,\zeta_n^{\pm1}]\oplus
\C[\zeta_1^{\pm1},\ldots,\zeta_n^{\pm1}]$.
Moreover, ${}_{n}X_{n-2}^{(i,j)}(\la_1,\cdots,\la_n)$ does 
not have a pole at $\la_{i,j}=0$. 
\end{lem}
\begin{proof}
The regularity at $\la_{i,j}=0$ is a consequence of 
the property \eqref{eq:odd}. 

Let $\hat{X}$ denote the expression \eqref{eq:X-pole}. 
We must show that the poles 
$[\la_{i,p}\pm1]=0$ ($1\le p\le i-1$),
$[\la_{j,p}\pm1]=0$ ($1\le p\le j-1, p\neq i$) 
contained in the product of $R$ matrices are spurious. 
We show $\hat{X}$ is regular at 
$[\la_{j,p}+1]=0$ with $1\le p\le i-1$ and 
$[\la_{i,p}+1]=0$ with $1\le p\le i-1$. 
The other cases can be treated in a similar manner. 

Using the Yang-Baxter relation \eqref{eq:YBE}
we can move $L_j$ to the right 
and bring it to the position next to $L_p$,   
so that $\hat{X}$ becomes as follows. 
\be
&&\hat{X}(u)=
R_{i,i-1}\cdots R_{i,1}\cdot R_{j,j-1}
\cdots 
\widehat{R_{j,i}}\cdots R_{j,p+1}
\cdot 
R_{\overline{i-1},\bar{i}}\cdots R_{\bar{1},\bar{i}}
\\
&&\quad\times
\Tr_{\la_{i,j}}\left(L_{\bar{1}}\cdots\widehat{L_{\bar{i}}}\cdots
L_{\bar{n}}\cdot L_{n}\cdots \widehat{L_{j}}\cdots
\widehat{L_{i}}\cdots L_{p+1}
\underline{L_jL_pR_{j,p}}L_{p-1}\cdots L_1\right)
\\
&&\quad\times
R_{j,p-1}\cdots R_{j,1}\cdot R_{\overline{j-1},\bar{j}}
\cdots \widehat{R_{\bar{i},\bar{j}}}\cdots 
\underline{R_{\bar{p},\bar{j}}}\cdots R_{\bar{1},\bar{j}}
\bigl(s_{i,\bar{j}}s_{\bar{i},j}
u_{1,\cdots,\hat{i},\cdots,\hat{j},\cdots,n,
\bar{n},\cdots,\hat{\bar{j}},\cdots,\hat{\bar{i}},\cdots,\bar{1}}\bigr).
\en
Here we have set 
$L_{\bar{k}}=L_{\bar{k}}((\la_i+\la_j)/2-\la_k-1)$, 
$L_k=L_k((\la_i+\la_j)/2-\la_k)$, 
$R_{k,l}=R_{k,l}(\la_{k,l})$ and 
$R_{\bar{k},\bar{l}}=R_{\bar{k},\bar{l}}(\la_{k,l})$.
The possible pole at $[\la_{j,p}+1]=0$ comes only from the 
underlined part,  
\be
L_j\Bigl(\frac{\la_{i,j}}{2}\Bigr)
L_p\Bigl(\frac{\la_{i,j}}{2}+\la_{j,p}\Bigr)
R_{j,p}(\la_{j,p})
\quad \mbox{ and }\quad  
R_{\bar{p},\bar{j}}\Bigl(\la_{p,j}\Bigr). 
\en
Because of \eqref{eq:q-det-modified} and 
\eqref{eq:L-parity}, the pole is in fact absent. 

Likewise, we move $L_{\bar{j}}$ to the left, and 
use cyclicity \eqref{eq:tr1} and crossing symmetry \eqref{eq:crossing}
to change it to $L_{i}$. Bringing the 
latter further to the left, we obtain 
\be
&&-\hat{X}(u)=
R_{i,i-1}\cdots R_{i,p+1}
\cdot 
R_{\overline{i-1},\bar{i}}\cdots 
\underline{R_{\bar{p},\bar{i}}}\cdots
R_{\bar{1},\bar{i}}\cdot
R_{\overline{j-1},\bar{j}}\cdots 
\widehat{R_{\overline{i},\bar{j}}}
\cdots R_{\bar{1},\bar{j}}
\\
&&\quad\times
\Tr_{\la_{i,j}}\left(L_{\bar{1}}\cdots\widehat{L_{\bar{i}}}\cdots
\widehat{L_{\bar{j}}}\cdots
L_{\bar{n}}\cdot L_{n}\cdots \widehat{L_{i}}\cdots L_{p+1}
\underline{L_iL_pR_{i,p}}L_{p-1}\cdots L_1\right)
\\
&&\quad\times
R_{i,p-1}\cdots R_{i,1}\cdot 
R_{j,j-1} \cdots \widehat{R_{j,i}}\cdots R_{j,1}
\bigl(s_{i,\bar{j}}s_{\bar{i},j}
u_{1,\cdots,\hat{i},\cdots,\hat{j},\cdots,n,
\bar{n},\cdots,\hat{\bar{j}},\cdots,\hat{\bar{i}},\cdots,\bar{1}}\bigr),
\en
where the underlined part is 
\be
R_{\bar{p},\bar{i}}(\la_{p,i})
\quad \mbox{ and }\quad  
L_i\Bigl(\frac{\la_{j,i}}{2}\Bigr)
L_p\Bigl(\frac{\la_{j,i}}{2}+\la_{i,p}\Bigr)
R_{i,p}(\la_{i,p}). 
\en
Hence $[\la_{i,p}+1]=0$ is not a pole for the same reason as above. 
\end{proof}

\section{Proof of  exchange relation}\label{sec:exchange}

Let us verify the exchange relation. 

In the case $(i,j)=(k,k+1)=(1,2)$, 
\eqref{eq:X-Rsym} is derived by noting the Yang-Baxter relation
\bea
\Rc_{1,2}(\la_{1,2})L_2(\la-\la_2)L_1(\la-\la_1)
=
L_2(\la-\la_1)L_1(\la-\la_2)\Rc_{1,2}(\la_{1,2})
\label{RLL}
\ena
and that  
\be
&&
\Rc_{1,2}(\la_{1,2})\Rc_{\bar2,\bar1}(\la_{2,1})
L_{2}\bigl(\frac{\la_{1,2}}{2}\bigr)
L_{\bar 2}\bigl(\frac{\la_{1,2}}{2}-1\bigr)s_{1,\bar2}s_{\bar1,2}
\\
&=&
-\Rc_{1,2}(\la_{1,2})\Rc_{\bar2,\bar1}(\la_{2,1})
L_{2}\bigl(\frac{\la_{1,2}}{2}\bigr)
L_{1}\bigl(\frac{\la_{2,1}}{2}\bigr)s_{1,\bar2}s_{\bar1,2}
\\
&=&
-L_{2}\bigl(\frac{\la_{2,1}}{2}\bigr)
L_{1}\bigl(\frac{\la_{1,2}}{2}\bigr)
\Rc_{1,2}(\la_{1,2})\Rc_{\bar2,\bar1}(\la_{2,1})
s_{1,\bar2}s_{\bar1,2}
\\
&=&
L_{2}\bigl(\frac{\la_{2,1}}{2}\bigr)
L_{\bar 2}\bigl(\frac{\la_{2,1}}{2}-1\bigr)s_{1,\bar2}s_{\bar1,2}.
\en
In the last line we used the unitarity
\[
R_{a,b}(\la)R_{b,a}(-\la)=\id
\]
and crossing symmetry
\[
R_{a,b}(\la)s_{a,c}=-R_{b,c}(-\la-1)s_{a,c}
\]
for the $R$ matrix.

The rest of the assertion is an easy consequence of the Yang-Baxter relation
\eqref{RLL}.

\section{Proof of the 
first difference equation}\label{sec:first}

In this section we prove 
\eqref{eq:X-diff1} and \eqref{eq:X-diff2}. 
We note that the symmetry 
\be
&&
A^{(1)}_n(\cdots,\la_{j+1},\la_j,\cdots)
\Rc_{j,j+1}(\la_{j,j+1})\Rc_{\overline{j+1}\bar{j}}(\la_{j+1,j})
\\
&&\quad =
\Rc_{j,j+1}(\la_{j,j+1})\Rc_{\overline{j+1}\bar{j}}(\la_{j+1,j})
A^{(1)}_n(\cdots,\la_{j},\la_{j+1},\cdots)\quad(j\geq2)\label{SYMA1}
\en
allows us to restrict to $j=2$ 
for \eqref{eq:X-diff1},  
and to $(i,j)=(2,3)$ for \eqref{eq:X-diff2}
without loss of generality.

Observe also that the relations \eqref{eq:X-Rsym}--\eqref{eq:X-red}
are equalities 
between rational functions in $\zeta_j$. 
It is enough if we prove them when 
one of the parameter differences 
$\la_{i,j}$ is a sufficiently large integer, 
because of the following Lemma.  

\begin{lem}\label{lem:integer}
Let $q$ be a non-zero complex number such that $q^n\neq 1$ 
for any positive integer $n$, and let $f(\la,\zeta)\in\C[\la,\zeta]$.  
If $f(k,q^k)=0$ for any integer $k>N$ for some $N$, then 
$f(\la,\zeta)=0$. 
\end{lem}
\begin{proof}
Write $f(\la,\zeta)=\sum_{m=0}^Mf_m(\zeta)\la^m$, $f_m(\zeta)\in\C[\zeta]$. 
We show $f=0$ by induction on $M$ and $\deg f_M$. 
The case $M=0$ is obvious. 
Suppose the assertion is proved for smaller values of $M$, or 
the same $M$ with  $\deg f_M<l$. Let us prove the case $\deg f_M\le l$. 
Consider $\tilde{f}(\la,\zeta)=f(\la+1,q\zeta)-q^lf(\la,\zeta)$.   
By the induction hypothesis, we have $\tilde{f}=0$. 
Let $T$ be the linear map on 
$W=\oplus_{m=0}^M\oplus_{n=0}^N\C\la^m\zeta^n$ 
($N=\max\deg f_m$)  given by 
$(Tg)(\la,\zeta)=g(\la+1,q\zeta)$.   
We have then $Tf=q^lf$.  
Since $q$ is not a root of unity, 
$W$ is a direct sum of generalized eigenspaces 
$W_s=\oplus_{m=0}^M\C\la^m\zeta^s$ of $T$ with the 
eigenvalue $q^s$ ($0\le s\le N$). 
Each eigenspace is one-dimensional because 
$T\bigl|_{W_s}$ is represented by a matrix with one Jordan block. 
Hence $Tf=q^lf$ implies $f=a\zeta^l$ with some $a\in \C$. 
It is then clear that $f=0$. 
\end{proof}

We prove \eqref{eq:X-diff1} for $(i,j)=(1,2)$,  
assuming $\la_{2,1}=k+1 \in\Z_{\ge 2}$. 

We use an auxiliary space $V_a^{(k)}$ of dimension $k+1$ and set 
\[
\pi^{(k)}_a=\pi^{(k)}\otimes\id_{V^{\otimes2n}}
~:~U_q(\slt)\otimes {\rm End}(V^{\otimes2n})
\longrightarrow
{\rm End}(V^{(k)}_a)\otimes
{\rm End}(V^{\otimes2n}).
\]
Since
\be
&&
-\prod_{p=2}^n[\la_{1,p}-1][\la_{1,p}+1]\cdot 
A^{(1)}_n(\la_1,\cdots,\la_n)
=\tr_{V^{(1)}_a}
\left(\pi^{(1)}_a(T^{[1]}_n(\la_1))P_{a,\bar1}\right)P_{1,\bar 1},
\en
the identity to be proved reads
\bea
&&\prod_{p=2}^n[\la_{2,p}-k-1][\la_{2,p}-k]\cdot
\tr_{V^{(k+1)}_c}
\left(\pi^{(k+1)}_c(T^{[1]}_n\bigl(\la_2-\frac{k+2}{2}\bigr)\right)
s_{1,\bar2}s_{\bar1,2} 
\label{eq:first}
\\
&&=\tr_{V^{(1)}_a}
\Bigl(\pi^{(1)}_a\bigl(T_n^{[1]}(\la_2-k-1)\bigr)P_{a,\bar1}\Bigr)P_{1,\bar 1}
\tr_{V_b^{(k)}}
\Bigl(\pi^{(k)}_b\bigl(T_n^{[1]}(\la_2-\frac{k+1}{2})\bigr)\Bigr)
s_{1,\bar2}s_{\bar1,2}. 
\nn
\ena
First, we show that the operator $P_{1,\bar 1}$ can be replaced
by $1$. Since $P_{1,\bar 1}=1+r_{1,\bar1}(-1)$, it is enough to show that
\begin{equation}
r_{1,\bar1}(-1)
\tr_{V_b^{(k)}}
\Bigl(\pi^{(k)}_b\bigl(T_n^{[1]}(\la_2-\frac{k+1}{2})\bigr)\Bigr)
s_{1,\bar2}s_{\bar1,2}=0.\label{REMOVEP}
\end{equation}
By using crossing symmetry (see \eqref{eq:crossing})
and cyclicity of the trace \eqref{eq:tr1}, we see that the
above expression contains
\[
r^{(k,1)}_{b,2}(\frac{\la_{1,2}}2)r^{(k,1)}_{b,\bar2}(\frac{\la_{1,2}}2-1)
r_{2,\bar2}(-1)
\]
inside the trace. Then, using \eqref{eq:q-det}, we obtain
\eqref{REMOVEP}.

Rewrite the right hand side of \eqref{eq:first} as 
\be
&&\tr_{V_a^{(1)}\otimes V_b^{(k)}}
\Bigl(AB\Bigr)s_{1,\bar2}s_{\bar1,2}
=
\tr_{V_a^{(1)}\otimes V_b^{(k)}}
\Bigl(BA\Bigr)s_{1,\bar2}s_{\bar1,2},
\en
where
\be
&&A=r_{a,\bar2}(-k-2)r_{b,\bar2}^{(k,1)}(-\frac{k+3}{2})
\pi^{(1)}_a(T_n^{[1,2]}(\la_2-k-1))
\pi^{(k)}_b(T^{[1,2]}_n(\la_2-\frac{k+1}{2})),
\\
&&
B=r_{a,2}(-k-1)r_{b,2}^{(k,1)}(-\frac{k+1}{2})P_{a,\bar1}.
\en
Here we used
\[
T^{[1,2]}_n(\la)=L_{\bar3}(\la-\la_3-1)\cdots L_{\bar n}(\la-\la_n-1)
L_n(\la-\la_n)\cdots L_3(\la-\la_3).
\]

The space $V^{(k)}$ can be identified with the subspace 
of $V^{\otimes k}$ consisting of completely symmetric tensors,
\bea
V^{(k)}\simeq 
\bigcap_{i=1}^{k-1}
\Ker r_{b_{i+1},b_i}(-1)\subset 
V_{b_k}^{(1)}\otimes \cdots\otimes V_{b_1}^{(1)},
\label{eq:embed}
\ena
and the operator $r^{(k,1)}_{b,2}(\la)$ acting on $V^{(k)}_b\otimes V^{(1)}_2$
is identified as
\[
r^{(k,1)}_{b,2}(\la)\prod_{j=0}^{k-2}[\la+\frac{k-1}2-j]
=r_{b_k,2}(\la-\frac{k-1}2)\cdots r_{b_1,2}(\la+\frac{k-1}2).
\]
The factor in the left hand side is non-zero for
$\la=\la_{1,2}/2=-(k+1)/2$.
Let us show 
\be
B\left(V^{(1)}\otimes V^{(k)}\right)\subset V^{(k+1)}. 
\en
It suffices to check that
\be
r_{a,b_k}(-1)Bs_{\bar1,2}=0. 
\en
Let us prove it. 
{}From the Yang-Baxter equation we have 
\begin{eqnarray*}
r_{a, b_{k}}(-1)Bs_{\bar{1}, 2}&=&
r_{b_{k}, 2}(-k) r_{a, 2}(-k-1) r_{b_{k-1}, 2}(-k+1) \cdots 
r_{b_{2}, 2}(-2) \\ 
&\times& 
r_{a, b_{k}}(-1) r_{b_{1}, 2}(-1) 
P_{a, \bar{1}} s_{\bar{1}, 2}.
\end{eqnarray*}
By using crossing symmetry repeatedly, we find 
\begin{eqnarray*}
r_{b_{1}, 2}(-1)P_{a, \bar{1}}s_{\bar{1}, 2}&=&
r_{b_{1}, 2}(-1)r_{a, \bar{1}}(0)s_{\bar{1}, 2} \\ 
&=&
{}-r_{b_{1}, 2}(-1)r_{2, a}(-1)s_{\bar{1}, 2}=
r_{a, b_{1}}(0)r_{2, a}(-1)s_{\bar{1}, 2}.  
\end{eqnarray*}
Thus we get 
\begin{eqnarray*}
r_{a, b_{k}}(-1) r_{b_{1}, 2}(-1) 
P_{a, \bar{1}} s_{\bar{1}, 2}&=&
r_{a, b_{k}}(-1)r_{a, b_{1}}(0)r_{2, a}(-1)s_{\bar{1}, 2} \\ 
&=&r_{a, b_{1}}(0)r_{b_{1}, b_{k}}(-1)r_{2, a}(-1)s_{\bar{1}, 2}.  
\end{eqnarray*}
This acts as $0$ on $V^{(1)}\otimes V^{(k)}$.

Therefore the trace of $BA$ is unchanged if we replace
$V^{(1)}\otimes V^{(k)}$ by $V^{(k+1)}$. 
Using crossing symmetry we 
rewrite the right hand side of \eqref{eq:first} as 
\bea
&&
\tr_{V^{(k+1)}}\Bigl(BA\Bigr)s_{1,\bar2}s_{\bar1,2}
=-
\tr_{V^{(k+1)}}\Bigl(A
r_{a,2}(-k-1)P_{a,\bar 1}
r_{b,\bar1}^{(k,1)}\bigl(\frac{k-1}{2}\bigr)
\Bigr)s_{1,\bar2}s_{\bar1,2}.
\label{eq:P1a}
\ena
The space $V^{(k+1)}$ can also be identified with
$\Im r_{b,a}^{(k,1)}\bigl(\frac{k+1}{2}\bigr)
\subset V^{(k)}\otimes V^{(1)}$. 
On the other hand, \eqref{eq:q-det} implies that
\be
\cP^-_{a,\bar1}
r^{(k,1)}_{b,\bar1}\bigl(\frac{k-1}{2}\bigr)
r_{b,a}^{(k,1)}\bigl(\frac{k+1}{2}\bigr)=0
\en
holds on $V^{(k)}\otimes V^{(1)}$. 
Hence in the right hand side of 
\eqref{eq:P1a} one can drop $P_{a,\bar1}$. 

Using the fusion relation 
\be
r^{(1,1)}_{b,a}\bigl(\la-\frac{k}{2}\bigr)
r^{(k,1)}_{(b_1,\cdots,b_k),a}
(\la+\frac{1}{2})\Bigl|_{V^{(k+1)}}
=\left[\la-\frac{k}{2}+1\right]\cdot 
r^{(k+1,1)}_{(b,b_1,\cdots,b_{k}),a}(\la),
\label{eq:fusion}
\en
we obtain \eqref{eq:first}.

\section{Proof of the second difference equation}\label{sec:second}

We prove \eqref{eq:X-diff2} with $(i,j)=(2,3)$ and $k=1$.  
Other cases follow from \eqref{eq:X-Rsym} and the following equalities.
\be
&&A^{(j)}_n(\ldots,\la_{k+1},\la_k,\ldots)
\Rc_{k,k+1}(\la_{k,k+1})\Rc_{\overline{k+1},\overline{k}}(\la_{k+1,k})\nn\\
&&=
\begin{cases}
\Rc_{k,k+1}(\la_{k,k+1})\Rc_{\overline{k+1},\overline{k}}(\la_{k+1,k})
A_n^{(j)}(\ldots,\la_k,\la_{k+1},\ldots)&\hbox{ if }j\not=k,k+1;\\
\Rc_{k,k+1}(\la_{k,k+1}-1)\Rc_{\overline{k+1},\overline{k}}(\la_{k+1,k}+1)
A_n^{(k)}(\ldots,\la_k,\la_{k+1},\ldots)&\hbox{ if }j=k+1;\\
\Rc_{k,k+1}(\la_{k,k+1}+1)\Rc_{\overline{k+1},\overline{k}}(\la_{k+1,k}-1)
A_n^{(k+1)}(\ldots,\la_k,\la_{k+1},\ldots)&\hbox{ if }j=k.
\end{cases}
\en

The relation to be shown reads as follows:
\begin{eqnarray}
&&{}_{n}X_{n-2}^{(2,3)}(\la_1-1,\la_2,\cdots,\la_n) 
t^{[1]}_{n-2,\bar 1}(\la_1; \la_{4}, \ldots , \la_{n})P_{1,\bar1}
\label{eq:second}\\
&&=R_{\bar1,\bar2}(\la_{1,2}-1)R_{\bar1,\bar3}(\la_{1,3}-1)
t^{[1,2,3]}_{n,\bar 1}(\la_1; \la_{4}, \ldots , \la_{n})
R_{\bar1,3}(\la_{1,3})R_{\bar1,2}(\la_{1,2})
P_{1,\bar1}
\nn\\
&&\times
{}_{n}X_{n-2}^{(2,3)}(\la_1,\la_2,\cdots,\la_n). 
\nn 
\end{eqnarray}
Here $t^{[1]}_{n-2,\bar{1}}$ and $t^{[1,2,3]}_{n,\bar 1}$ are given in 
\eqref{eq:t_a}. 

{}From \eqref{eq:X-diff1}, we have
\be
{}_{n}X_{n-2}^{(2,3)}(\la_1,\la_2-1,\cdots,\la_n)
=-A^{(2)}_n(\la_1,\cdots,\la_n)\circ
{}_{n}X_{n-2}^{(2,3)}(\la_1,\la_2,\cdots,\la_n).
\en
Let $Z(\la_1,\cdots,\la_n)$ denote the left hand side minus the 
right hand side of \eqref{eq:second}. 
Then 
\be
Z(\la_1,\la_2-1,\cdots,\la_n)
=-A^{(2)}_n(\la_1-1,\cdots,\la_n)Z(\la_1,\la_2,\cdots,\la_n). 
\en
The matrix $A^{(2)}(\la_1-1,\cdots,\la_n)$ does not have a pole at
$\la_2=\la_3-k$ with $k\in\Z$, $k\ge 2$. 
To show $Z=0$, it is therefore 
enough to check it for $\la_2=\la_3-2$. 

The verification is straightforward with the aid of 
\bea
&&
{}_{n}X_{n-2}^{(2,3)}(\la_1,\la_3-2,\la_3, \ldots , \la_{n})(u) 
\label{eq:X23la=2}
\\
&& {}=
(-1)^n\frac{1}{[2]}r_{\bar2,\bar3}(-2)R_{2,1}(\la_{3,1}-2)
R_{\bar1,\bar3}(\la_{1,3})
\nn\\ 
&& {}\times 
t^{[1,2,3]}_{n,\bar 2}(\la_3-1; \la_{4}, \ldots , \la_{n}) 
(s_{2,\bar3}s_{\bar2,3}
u_{1, 4, \ldots , n, \bar{n}, \ldots , \bar{4}, \bar{1}}). 
\nn
\ena

\section{Proof of commutativity}\label{sec:X-comm}

{}From the exchange relation \eqref{eq:X-Rsym} 
it suffices to prove the case of $(i, j, k, l)=(1, 2, 3, 4)$. 
Let 
$Z'(\la_1,\cdots,\la_n)$
stand for the difference between the left hand side and the
right hand side of \eqref{eq:X-comm}.
{}From \eqref{eq:X-diff2},
we have
\begin{eqnarray*}
&&
Z'(\la_1-1, \la_{2}, \cdots,\la_n)=
-A^{(1)}_n(\la_1,\cdots,\la_n)Z'(\la_1, \la_{2}, \cdots,\la_n),
\\
&&
Z'(\cdots,\la_3-1,\cdots)=
-A^{(3)}_n(\la_1,\cdots,\la_n)Z'(\cdots,\la_3,\cdots). 
\end{eqnarray*}

Both $A^{(1)}_n(\la_1,\cdots,\la_n)$, 
$A^{(3)}_n(\la_1,\cdots,\la_n)$ 
do not have poles at $\la_{1,2}=-k$ nor $\la_{3,4}=-k$ for
$k\in\Z$, $k\ge 2$. 
Therefore, in order to show $Z'=0$, it suffices to verify this 
when $\la_{1,2}=\la_{3,4}=-2$. 
In this case, a lengthy but direct verification using \eqref{eq:Xla=2}
shows that both sides of \eqref{eq:X-comm} are equal to 
\begin{eqnarray*}
&&
-\frac1{[2]^{2}}r_{\bar1,\bar2}(-2)r_{\bar3,\bar4}(-2)
R_{\bar1,\bar4}(\la_{2,4}-2)
t^{[1,2,3,4]}_{n,\bar3}(\la_4-1; \la_{5}, \ldots , \la_{n}) \\ 
&& {}\times
t^{[1,2,3,4]}_{n,\bar1}(\la_2-1; \la_{5}, \ldots , \la_{n})
R_{\bar1,3}(\la_{2,4}+1)
s_{1,\bar2}s_{\bar1,2}s_{3,\bar4}s_{\bar3,4}. 
\end{eqnarray*}

\section{Proof of cancellation identity}\label{sec:X-last}

In this subsection, 
for $u\in V^{\otimes 2(n-1)}$ we abbreviate
$u_{1,\cdots,n-1,\overline{n-1},\cdots,\bar{1}}s_{n,\bar{n}}$ to 
$u\cdot s_{n,\bar{n}}$. Likewise 
for $u\in V^{\otimes 2(n-2)}$ we abbreviate
$u_{1,\cdots,n-2,\overline{n-2},\cdots,\bar{1}}
s_{n-1,\overline{n-1}}s_{n,\bar{n}}$ to 
$u\cdot s_{n-1,\overline{n-1}}s_{n,\bar{n}}$.
We set 
\be
&&Q_n(\zeta_1,\cdots,\zeta_n)
\\
&&\quad :=
4\sum_{j=2}^n
{}_{n}Y_{n-2}^{(1,j)}(\z_1,\cdots,\z_n)
(\bs_{n-2})
-\left(A^{(1)}_n(\la_1,\cdots,\la_n)^{-1}-1\right)
\bs_n.
\en

This is a rational function in $\zeta_1,\cdots,\zeta_n$.  
The following properties are easy to establish. 
\begin{eqnarray}
&&Q_n(\zeta_1/\xi,\cdots,\zeta_n/\xi)
=Q_n(\zeta_1,\cdots,\zeta_n) 
\qquad (\mbox{for any $\xi\in\C^{\times}$}),
\label{eq:homog}
\\
&&
\sum_{j=1}^n\left(\sigma_j^z+\sigma^z_{\overline{j}}\right) 
Q_n(\zeta_1,\cdots,\zeta_n)=0,
\label{eq:Q-wt0}
\\
&&P_{\overline j,\overline{j+1}}r_{\overline j,\overline{j+1}}(\la_{j,j+1})
Q_n(\cdots,\zeta_{j+1},\zeta_j,\cdots)
\label{eq:Q-Rsym}\\
&&=
P_{j,j+1}r_{j,j+1}(\la_{j,j+1})
Q_n(\cdots,\zeta_{j},\zeta_{j+1},\cdots)
\qquad (2\le j\le n-1), 
\nn
\\
&&
Q_n(\zeta_1,\cdots, \zeta_{n-1}, -\zeta_n)
=-\sigma_n^z\sigma_{\bar n}^z
Q_n(\zeta_1,\cdots, \zeta_{n-1}, \zeta_n).
\label{eq:Q-parity} 
\end{eqnarray}

Eq. \eqref{eq:Q-wt0} follows from \eqref{eq:tr2}.
Eq. \eqref{eq:Q-Rsym} follows from \eqref{SYMA1}, \eqref{eq:X-Rsym} and
\bea
\Rc_{12}(\la)\Rc_{\bar2,\bar1}(-\la)s_{1\bar1}s_{2\bar2}
=s_{1\bar1}s_{2\bar2}.\label{RRSS}
\ena
Eq. \eqref{eq:Q-parity} follows from Lemma \ref{lem:parity}.

\begin{lem}\label{lem:Q6}
As $\zeta_n^{\pm 1}\to\infty$, 
\be
Q_n(\zeta_1,\cdots,\zeta_n)
=Q_{n-1}(\zeta_1,\cdots,\zeta_{n-1})\cdot s_{n,\bar n}
+O(\zeta_n^{\mp1}).
\en
\end{lem}
\begin{proof}
Let us estimate the behavior of the difference
\bea
&&
Q_n(\zeta_1,\cdots,\zeta_n)
-Q_{n-1}(\zeta_1,\cdots,\zeta_{n-1})\cdot s_{n,\bar n}
\label{eq:Q-Q}
\\
&&
\quad
=
4\, {}_{n}Y^{(1,n)}_{n-2}(\z_1,\cdots,\z_n)
(\bs_{n-2})
\nn\\
&&\quad+
4\sum_{j=2}^{n-1}
\left({}_{n}Y^{(1,j)}_{n-2}(\z_1,\cdots,\z_n)(\bs_{n-2})
-{}_{n-1}Y^{(1,j)}_{n-3}(\z_1,\cdots,\z_{n-1})(\bs_{n-3})\cdot s_{n, \bar{n}} \right)
\nn\\
&&
\quad-
A^{(1)}_n(\la_1,\cdots,\la_n)^{-1}(\bs_{n})
{}+A^{(1)}_{n-1}(\la_1,\cdots,\la_{n-1})^{-1}(\bs_{n-1})\cdot s_{n, \bar{n}}.
\nn
\ena
Consider the behavior of the first term when $\zeta_n^{\pm 1}\to\infty$.
It is easy to estimate
$A^{(1)}_n(\la_1,\cdots,\la_n)^{-1}\bs_n$.
For ${}_{n}Y_{n-2}^{(1,n)}(\z_1,\cdots,\z_n)(\bs_{n-2})$,
we move $\la_n$ to the left end by using \eqref{eq:X-Rsym}.
The matrices $R_{n,j}(\la_{n,j})R_{\bar j,\bar n}(\la_{j,n})$
are regular, while $\pt(\la_{1,n}),p(\la_{1,n})$
behaves as $O(\zeta_n^{\mp 2})$.
We apply Lemma \ref{lem:power_of_trace}.
Because of the presence of $s_{1,\bar2}s_{\bar1,2}\bs_{n-2}$,
in the notation of Lemma \ref{lem:power_of_trace} we have
$\sum_{p=1}^n(\epsilon'_p+\bar\epsilon'_p)=0$. Therefore,
from $l=l'$ we obtain $l=-(\epsilon_1+\bar\epsilon'_1)$ and
$|l|\leq2$. Thus we have
\be
&&
{}_{n}\Ft^{(1,2)}_{n-2}(\z_n,\z_1,\cdots,\z_{n-1})
(\bs_{n-2})
=O(\zeta_n^{\pm 1}), 
\\
&&
{}_{n}\F^{(1,2)}_{n-2}(\z_n,\z_1,\cdots,\z_{n-1})
(\bs_{n-2})
=O(\zeta_n^{\pm 1}). 
\en

For the $j$-th term of the second sum, we have inside the trace 
\be
&&
\left(
\frac{1}{[\la_{1,n}][\la_{j,n}]}
L_{\bar n}\Bigl(\frac{\la_{1,n}+\la_{j,n}}{2}-1\Bigr)
L_{n}\Bigl(\frac{\la_{1,n}+\la_{j,n}}{2}\Bigr)
-1\right)
s_{n,\bar n}
\\
&&=
\left(
-
\frac{1}{[\la_{1,n}][\la_{j,n}]}
L_{\bar n}\Bigl(\frac{\la_{1,n}+\la_{j,n}}{2}-1\Bigr)
L_{\bar n}\Bigl(-\frac{\la_{1,n}+\la_{j,n}}{2}-1\Bigr)
-1\right)
s_{n,\bar n}.
\en
Inserting the formula \eqref{eq:L} of $L$, 
we see that this expression behaves as $O(\zeta_n^{\mp 1})$
as $\zeta_n^{\pm 1}\to\infty$. 
Likewise the third term of \eqref{eq:Q-Q} has the same behavior. 

This completes the proof. 
\end{proof}

\begin{lem}\label{lem:Q5}
The function $\prod_{j=2}^n[\la_{1,j}]\cdot Q_n(\zeta_1,\cdots,\zeta_n)$
belongs to $\C[\zeta_1^{\pm1},\ldots,\zeta_n^{\pm1}]$.
\end{lem}
\begin{proof}
The only possible poles are where one of the following 
factors vanish:
$[\la_{1,j}]$, $[\la_{1,j}\pm 1]$, 
$[\la_{1,j}-2]$ ($2\le j\le n$), 
$[\la_{i,j}]$ ($2\le i<j\le n$). 
In view of the parity property \eqref{eq:Q-parity}, 
it is enough to show that 
$\la_{1,j}\pm1=0$, $\la_{1,j}-2=0$ and 
$\la_{i,j}=0$ are not poles. 

At $\la_{1,j}=\pm 1$, only ${}_{n}Y^{(1,j)}_{n-2}$ can have a pole. 
Since
\be
\res_{\la=\pm 1}\pt(\la)=\mp\res_{\la=\pm 1}p(\la),
\en
we have by Lemma \ref{lem:special}
\be
\res_{\la_{1,j}=\pm 1}\({}_{n}Y^{(1,j)}_{n-2}(\z_1,\cdots,\z_n)\)
&=&const\times {}_{n}X^{(1,j)}_{n-2}(\la_1,\cdots,\la_n)\Bigl|_{\la_{1,j}=\pm 1}
\\
&=&0.
\en
The last line follows from Lemma \ref{lem:pole}.

At $\la_{i,j}=0$ ($2\le i<j\le n$), the contributing terms are
${}_{n}Y^{(1,i)}_{n-2}(\bs_{n-2})$ and ${}_{n}Y^{(1,j)}_{n-2}(\bs_{n-2})$. 
It is enough to compare
${}_{n}X^{(1,i)}_{n-2}(\bs_{n-2})$ and ${}_{n}X^{(1,j)}_{n-2}(\bs_{n-2})$.
It is easy to check that their residues cancel each other
by using $P_{i+1,i}r_{i+1,i}(0)=1$, \eqref{RLL} and \eqref{RRSS} .

Consider $\la_{1,j}=2$. Using 
\be
[\la-2]\,\pt(\la)\Bigl|_{\la=2}=\frac{1}{2},
\quad 
[\la-2]\,p(\la)\Bigl|_{\la=2}=-\frac{1}{4},
\en
we find
\begin{eqnarray*}
&&
\prod_{p=2}^n[\la_{1,p}][\la_{1,p}-2]\cdot
4\, {}_{n}Y^{(1,j)}_{n-2}(\z_1,\cdots,\z_n)\Bigl|_{\la_{1,j}=2}(\bs_{n-2})
\\
&&{}=
-[2]\prod_{p=2 \atop p \not=j}^{n}
[\lambda_{j,p}+2][\lambda_{j, p}]\cdot {}_{n}X_{n-2}^{(1, j)}
(\lambda_{1}, \ldots , \lambda_{n})\Bigl|_{\la_{1,j}=2}(\bs_{n-2})\\
&&{}=
-R_{j,j-1}(\la_{j,j-1})\cdots R_{j,2}(\la_{j,2})
R_{\overline{j-1},\bar j}(\la_{j-1,j})\cdots 
R_{\bar 2,\bar j}(\la_{2,j})
\\
&&\times 
\Tr_2\left(L_{\bar j}(0)T^{[1,j]}_n(\la_1-1)L_{j}(1)\right)
s_{1\bar j}s_{\bar 1,j}\prod_{p(\neq 1,j)}s_{p,\bar p}, 
\end{eqnarray*}
where 
\begin{eqnarray*}
T^{[1, j]}_n(\lambda_{1}-1)&=&L_{\bar{2}}(\lambda_{1,2}-2) \cdots 
\widehat{L_{\bar{j}}(\lambda_{1, j}-2)} \cdots 
L_{\bar{n}}(\lambda_{1, n}-2) \\
&\times&  
L_{n}(\lambda_{1, n}-1) \cdots 
\widehat{L_{j}(\lambda_{1, j}-1)} \cdots 
L_{2}(\lambda_{1, 2}-1).
\end{eqnarray*}
Under the ${\rm Tr}_2$, this expression becomes
\bea
&&(-1)^jr_{2,1}(\la_{2,1})\cdots r_{n,1}(\la_{n,1})
r_{\bar n,1}(\la_{n,1}+1)\cdots 
r_{\overline{j+1},1}(\la_{j+1,1}+1)
\label{eq:la=2}\\
&&\quad 
\times
r_{\bar 1,\bar 2}(\la_{j,2})\cdots 
r_{\bar 1,\overline{j-1}}(\la_{j,j-1})
s_{1,\bar j}s_{\bar 1,j}\prod_{p(\neq 1,j)}s_{p,\bar p}.\nonumber
\ena
Inserting
\be
-s_{1,\bar j}s_{\bar 1,j}
=r_{\bar j,1}(-1)s_{1,\bar 1}s_{j,\bar j}
\en
the last line is rewritten as 
\be
(-1)^{j-1}r_{\bar j,1}(\la_{j,1}+1)
r_{\overline{j-1},1}(\la_{j-1,1}+1)
\cdots
r_{\overline{2},1}(\la_{2,1}+1)\prod_{p=1}^ns_{p,\bar p}.
\en
Hence \eqref{eq:la=2} becomes
\be
&&\prod_{p=2}^n[\la_{1,p}][\la_{1,p}-2]
\cdot A^{(1)}_n(\la_1,\cdots,\la_n)^{-1}
(\bs_n)
\Bigl|_{\la_{1,j}=2}\\
&&\qquad=-r_{2,1}(\la_{2,1})\cdots r_{n,1}(\la_{n,1})
r_{\bar n,1}(\la_{n,1}+1)\cdots r_{\bar2,1}(\la_{2,1}+1)\bs_n
\en
as desired. 
\end{proof}

\begin{lem}\label{lem:Q7}
\be
\cP^-_{\overline{n-1},\overline{n}}
Q_n(\zeta_1,\cdots,\zeta_{n-1}, q^{-1}\zeta_{n-1})
=\cP^-_{\overline{n-1},\overline{n}}
\left(Q_{n-2}(\zeta_1,\cdots,\zeta_{n-2})\cdot
s_{n-1,\overline{n-1}}s_{n,\overline{n}}\right).
\en
\end{lem}
\begin{proof}
Consider
\bea
\cP^-_{\overline{n-1},\bar n}\circ 
{}_{n}X^{(1,j)}_{n-2}(\la_1,\cdots,\la_n)
(\bs_{n-2}).
\label{eq:PX}
\ena
For $2\le j\le n-2$, we have inside the trace the combination
\be
&&\cP^-_{\overline{n-1},\bar n}
L_{\overline{n-1}}\Bigl(\frac{\la_1+\la_j}{2}-\la_{n-1}-1\Bigr)
L_{\overline{n}}\Bigl(\frac{\la_1+\la_j}{2}-\la_{n}-1\Bigr)
\\
&&\times
L_{n}\Bigl(\frac{\la_1+\la_j}{2}-\la_{n}\Bigr)
L_{n-1}\Bigl(\frac{\la_1+\la_j}{2}-\la_{n-1}\Bigr)
\left(s_{n-1,\overline{n-1}}s_{n,\bar n}\right).
\en
{}From \eqref{eq:q-det-modified} and 
$\mathcal{P}_{\overline{n-1}, \overline{n}}^{-}s_{n-1, \overline{n-1}}
s_{n, \bar{n}}=
\mathcal{P}_{n-1, n}^{-}s_{n-1, \overline{n-1}}
s_{n, \bar{n}}$, 
this expression when specialized to $\la_n=\la_{n-1}-1$ 
reduces to
\be
[\la_{1,n-1}][\la_{j,n-1}]
[\la_{1,n-1}+1][\la_{j,n-1}+1]
\cP^-_{n-1, n}\left(s_{n-1,\overline{n-1}}s_{n,\bar n}\right).
\en
Hence \eqref{eq:PX} becomes 
\be
\cP^-_{\overline{n-1},\bar n}
\left( 
{}_{n-2}X^{(1,j)}_{n-4}(\la_1,\cdots,\la_{n-2})
(\bs_{n-4})\cdot 
s_{n-1, \overline{n-1}}s_{n, \bar{n}}
\right)
\quad (2\le j\le n-2).  
\en
Let us show that \eqref{eq:PX} vanishes for $j=n-1,n$. 
For $j=n-1$, this is due to the factor
\be
\cP^-_{\overline{n-1},\bar n}
L_{\overline{n-1}}\Bigl(\frac{\la_{1,n-1}}{2}-1\Bigr)
L_{\overline{n}}\Bigl(\frac{\la_{1,n-1}}{2}\Bigr)
\en
inside the trace $\Tr_{\la_{1,n-1}}$. 

In the case $j=n$, we have inside $\Tr_{\la_{1,n}}$ 
\begin{eqnarray*}
&&
\cP^-_{\overline{n-1},\bar n}
L_{\overline{n-1}}
\Bigl(\frac{\la_{1,n}}{2}+\la_{n,n-1}-1\Bigr)
L_{\overline{n}}\Bigl(\frac{\la_{1,n}}{2}-1\Bigr)
\frac{r_{\overline{n-1},\overline n}(\la_{n-1,n})}{[\la_{n-1,n}+1]} 
\\
&&\times
L_{n}\Bigl(\frac{\la_{1,n}}{2}\Bigr)
L_{n-1}\Bigl(\frac{\la_{1,n}}{2}+\la_{n,n-1}\Bigr)
\frac{r_{n,n-1}(\la_{n,n-1})}{[\la_{n,n-1}+1]}. 
\end{eqnarray*}
At $\la_{n-1}=\la_n+1$, 
the first product vanishes and the second product is regular. 

By a similar calculation we find 
\begin{eqnarray*}
&& 
\cP^-_{\overline{n-1},\bar n}
A^{(1)}_n(\la_1,\cdots,\la_n)^{-1}
(\bs_n)
\Bigl|_{\la_n=\la_{n-1}-1} \\ 
&& {}=
\mathcal{P}_{\overline{n-1}, \overline{n}}^{-}
\left( 
A_{n-2}^{(1)}(\lambda_{1}, \cdots , \lambda_{n-2})^{-1}(\bs_{n-2})\cdot 
s_{n-1, \overline{n-1}}s_{n, \bar{n}} 
\right). 
\end{eqnarray*}
The assertion follows from these.
\end{proof}
\medskip

Let us now prove  \eqref{eq:X-last}.

Lemma \ref{lem:Q5} implies that 
\be
u_n=\prod_{j=2}^n(\zeta_j/\zeta_1-\zeta_1/\zeta_j)
\cdot Q_n(\zeta_1,\cdots,\zeta_n)
\en
has no poles in 
$\zeta_{j} \in \mathbb{C}^{\times}$, ($j=1, \ldots , n$).
It is evident that $Q_0=Q_1=0$. 
Suppose $n\ge 2$, and assume
by induction that $Q_{n-2}=Q_{n-1}=0$. 
It follows from \eqref{eq:Q-Rsym} 
and Lemma \ref{lem:Q6}
that $u_n$ is regular at 
$\zeta_j^{\pm 1}\to\infty$ for $j=2,\cdots,n$.  
Along with \eqref{eq:homog}, we conclude that 
$u_n$ is independent of $\zeta_1,\cdots,\zeta_n$. 

It is convenient to pass from 
vectors in $V^{\otimes 2n}$ to matrices via the isomorphism 
\be
&&
V^{\otimes 2n}\overset{\sim}{\longrightarrow }
\End\left(V^{\otimes n}\right),
\\
&&v_{\epsilon_1}\otimes \cdots \otimes 
v_{\epsilon_n}\otimes v_{\overline{\epsilon}_n}
\otimes\cdots\otimes v_{\overline{\epsilon}_1}
\mapsto 
\prod_{p=1}^n(-{\overline{\epsilon}}_p)\cdot
E_{\epsilon_1,-{\overline{\epsilon}}_1}
\otimes\cdots\otimes  
E_{\epsilon_n,-{\overline{\epsilon}}_n}, 
\en
where $E_{\epsilon,\overline{\epsilon}}
=
\left(\delta_{\epsilon,a}\delta_{\overline{\epsilon},b}\right)
_{a,b}$ 
denotes the matrix unit. It induces the mapping
$\End(V^{\otimes 2n})\overset{\sim}{\longrightarrow }
\End(\End\left(V^{\otimes n}\right))$. In particular,
$\left(E_{\epsilon,\epsilon'}\right)_j$ in the former corresponds to the left
multiplication by $\left(E_{\epsilon,\epsilon'}\right)_j$ in the latter,
and $\left(E_{\epsilon,\epsilon'}\right)_{\overline j}$
in the former to the right multiplication by
\be
\begin{cases}
\left(E_{\mp,\mp}\right)_j&\hbox{ if }(\epsilon,\epsilon')=(\pm,\pm);\\
-\left(E_{\pm,\mp}\right)_j&\hbox{ if }(\epsilon,\epsilon')=(\pm,\mp).
\end{cases}
\en
in the latter.

Let $U_n\in\End(V^{\otimes n})$  
denote the constant matrix corresponding to the constant vector $u_n$. 
The properties \eqref{eq:Q-wt0}, \eqref{eq:Q-Rsym}, \eqref{eq:Q-parity}
of $Q_n$, Lemma \ref{lem:Q7} and the assumption of the induction
imply the following for $U_n$:
\bea
&&
[\sum_{j=1}^n\sigma^z_j,U_n]=0,
\label{eq:U1}
\\
&&[P_{j,j+1}r_{j,j+1}(\la_{j,j+1}),U_n]=0
\quad (2\le j\le n-1),
\label{eq:U2}
\\
&&\sigma_n^z U_n+U_n\sigma_n^z=0,
\label{eq:U3}
\\
&&
U_n \cP^-_{n-1,n}=0.
\label{eq:U4}
\ena

Letting $\zeta_j^{\pm1}\to\infty$ in \eqref{eq:U2}, 
we obtain in particular that 
\bea
[P_{j,j+1},U_n]=0
\quad (2\le j\le n-1).
\label{eq:U5}
\ena
Note that the linear space ${\rm End}(V^{\otimes n})$ has a basis 
\begin{eqnarray*}
\{ \tau_{1} \otimes \cdots \otimes \tau_{n} \, | \, 
\tau_{j}=\sigma^{+}, \sigma^{-}, \sigma^{z} \,\, \hbox{or} \,\, 1\} 
\end{eqnarray*}
Because of \eqref{eq:U3} and \eqref{eq:U5}, we can write
\be
U_n=\sum_{\epsilon_2,\cdots,\epsilon_n=\pm}
C_{\epsilon_2,\cdots,\epsilon_n}\otimes 
\sigma^{\epsilon_2}\otimes \cdots \otimes 
\sigma^{\epsilon_n}
\en
with some $C_{\epsilon_2,\cdots,\epsilon_n}\in\End(V)$. 
In view of \eqref{eq:U5}, they are 
symmetric with respect to $\epsilon_2,\cdots,\epsilon_n$. 
Since 
$(\sigma^{\pm} \otimes \sigma^{\pm})\cP^{-}=0$ and 
$(\sigma^+\otimes\sigma^-+\sigma^-\otimes\sigma^+)\cP^-\neq0$, 
\eqref{eq:U4} and \eqref{eq:U5} imply that $U_n$ has the form 
\be
U_n=\begin{cases}
A_+\otimes\sigma^+\otimes\cdots\otimes\sigma^+
+A_-\otimes\sigma^-\otimes\cdots\otimes\sigma^-& (n\ge 3)\\     
a_+\sigma^+\otimes\sigma^+
+a_-\sigma^-\otimes\sigma^- &(n=2) \\
\end{cases}
\en
with some $A_\pm\in \End(V)$, $a_\pm\in\C$.  
In either case, the zero weight property \eqref{eq:U1} enforces that $U_n=0$. 

Hence we have proved $Q_n=0$.

\section{Proof of  recurrence relation}\label{sec:Recurrence}

Finally let us prove \eqref{eq:X-red}. 

The case of $i=1$ follows from 
$\mathcal{P}_{1, \bar{1}}^{-}s_{1, \bar{2}}s_{\bar{1}, 2}=
\frac{1}{2}s_{1, \bar{1}}s_{2, \bar{2}}$ and \eqref{eq:q-det-modified}. 
Let us consider the case with $i \ge 2$. 

{}From the exchange relation \eqref{eq:X-Rsym} it suffices to 
prove the case of $(i, j)=(2, 3)$. 
Set
\be
Z''(\la_1,\cdots,\la_n)
=
{}_{n-1}\Pi_{n} \circ
{}_{n}X^{(2,3)}_{n-2}(\la_1,\cdots,\la_n)-
{}_{n-1}X^{(1,2)}_{n-3}(\la_2,\cdots,\la_n)\circ
{}_{n-3}\Pi_{n-2}.
\en
{}From \eqref{eq:X-diff1} for ${}_{n}X^{(2,3)}_{n-2}$ and using
\be
{}_{n-1}\Pi_{n}\circ
A^{(2)}_n(\la_1,\la_2,\ldots,\la_n)=
A^{(1)}_{n-1}(\la_2,\ldots,\la_n)\circ {}_{n-1}\Pi_{n}.
\en
we obtain
\be
Z''(\la_1,\la_2-1,\la_3,\cdots,\la_n)
=-
A^{(1)}_{n-1}(\la_2,\cdots,\la_n) 
\cdot Z''(\la_1,\la_2,\la_3,\cdots,\la_n).
\en
The initial conditions \eqref{eq:Xla=2} and \eqref{eq:X23la=2} imply that 
$Z''(\la_1,\la_3-2,\la_3,\cdots,\la_n)=0$. 
This proves $Z''(\la_1,\cdots,\la_n)=0$. 

\section{Another representation of $h_n$}
\label{sec:another}

There is an alternative way to write the formula for $h_n$,
which should be useful for further applications.
The main result of our study is given by \eqref{eq:def-hn}:
\be
&&h_n(\la_1,\cdots,\la_n)\\
&&=
\sum_{m=0}^{[n/2]}\frac{(-1)^m}{2^{n-2m}}\sum
\Omega_{K_1,(i_1,j_1)}\circ
\Omega_{K_2,(i_2,j_2)}\circ
\cdots \circ\Omega_{K_{m},(i_m,j_m)}
\left(\bs_{n-2m}\right),
\en
where $K_1$ is the standard set $\{1,\cdots ,n\}$,
$K_{p+1}=K_{p}\backslash \{i_p,j_p\}$, and $\Omega_{K,(i,j)}$ 
is related with ${}_m\Omega^{(k,l)}_{m-2}$
by \eqref{OMK}, $m=\sharp(K)$ being the cardinality of $K$.  
The main object of our formula 
$$
{}_n\Omega_{n-2}^{(i,j)}
(\la _1,\cdots ,\la _n)
\in \text{Hom}\(V^{\otimes 2(n-2)},V^{\otimes 2n}\)$$
is given by \eqref{eq:Xij}, \eqref{eq:X-F} and \eqref{eq:O-F}.

Recall ${}_{n-1}\Pi _{n}\in\Hom(V^{\otimes2n},V^{\otimes2(n-1)})$
given by \eqref{eq:def-Pi}.  
We use the notation 
${}_{n-2}\Pi _{n}={}_{n-2}\Pi _{n-1}\circ {}_{n-1}\Pi _{n}$, etc..
Let us introduce an operator
\bea
\qquad
\widehat{\Omega}_n^{(i,j)}(\la _1,\cdots ,\la _n)
=-4{}_n\Omega_{n-2}^{(i,j)}(\la _1,\cdots ,\la _n)
\circ
{}_{n-2}\Pi _{n}
\circ
\mathbb{R}_n^{(i,j)}(\la_1,\cdots,\la_n)^{-1}
\label{defOm}
\ena
where $\mathbb{R}_n^{(i,j)}$ is defined in \eqref{eq:complicated}.

Similarly to  \eqref{eq:O-F} we set
\begin{align}
\widehat{\Omega}^{(i,j)}_n(\la _1, \cdots ,\la _n)=
\ot(\la_i-\la_j)\widetilde{W}^{(i,j)}_n(\z_1,\cdots,\z_n)+
\omega(\la_i-\la_j)W^{(i,j)}_n(\z_1,\cdots,\z_n),\nn
\end{align}
where
\begin{align}
&\widetilde{W}^{(i,j)}_n(\z_1,\cdots,\z_n)=-4\widetilde{G}^{(i,j)}_n(\z_1,\cdots,\z_n)
\circ
{}_{n-2}\Pi _{n}
\circ
\mathbb{R}_n^{(i,j)}(\la_1,\cdots,\la_n)^{-1} \label{nonsense}
\\
&W^{(i,j)}_n(\z_1,\cdots,\z_n)=-4G^{(i,j)}_n(\z_1,\cdots,\z_n)
\circ
{}_{n-2}\Pi _{n}
\circ
\mathbb{R}_n^{(i,j)}(\la_1,\cdots,\la_n)^{-1}\nn
\end{align}
Notice that
$$
\widehat{\Omega }_n^{(i,j)}(\la _1,\cdots ,\la _n)
\in \End\(V^{\otimes 2n}\). 
$$
In what follows we shall often 
omit the arguments $\la_1,\cdots,\la_n$  
in $\widehat{\Omega}^{(i,j)}_n$, 
since they are always the same. 
\begin{lem}\label{lem:OmOm}
For $i<j$ and $k<l$ we have
\bea
&& \widehat{\Omega}_n^{(i,j)}(\la_1,\ldots,\la_n)
\circ {}_n\Omega_{n-2}^{(k,l)}(\la_1,\ldots,\la_n)
\nn\\
&&=\begin{cases}
0\qquad\qquad\qquad\qquad\qquad\qquad\qquad\qquad\qquad\qquad\quad\> 
\hbox{ if }\quad\{i,j\}\cap\{k,l\}\not=\emptyset;\\
-4{}_n\Omega_{n-2}^{(i,j)}(\la_1,\ldots,\la_n)
\circ {}_{n-2}\Omega_{n-4}^{(k',l')}(\la_1,\ldots,\widehat{\la_i},
\ldots,\widehat{\la_j},\ldots,\la_n)\\
\times {}_{n-4}\Pi _{n-2}\circ\mathbb{R}_{n-2}^{(i',j')}(\la_1,\ldots,\widehat{\la_k},
\ldots,\widehat{\la_l},\ldots,\la_n)^{-1}\qquad
\hbox{ if }\quad\{i,j\}\cap\{k,l\}=\emptyset.
\end{cases}\nn
\ena
Here $k'$ and $l'$ are the positions of $\la_k$ and $\la_l$ in
$(\la_1,\ldots,\widehat{\la_i},\ldots,\widehat{\la_j},\ldots,\la_n)$
and $i'$ and $j'$ are the positions of $\la_i$ and $\la_j$ in
$(\la_1,\ldots,\widehat{\la_k},\ldots,\widehat{\la_l},\ldots,\la_n)$.
\end{lem}
\begin{proof}
Using the exchange relation \eqref{eq:X-Rsym}
one finds
\be
&&\mathbb{R}_n^{(i,j)}(\la_,\ldots,\la_n)^{-1}
{}_n\Omega_{n-2}^{(k,l)}(\la _1,\cdots ,\la _n)\label{eq:ax}\\
&&=
{}_n\Omega_{n-2}^{(k'',l'')}(\la_i,\la_j,\la _1,
\ldots,\widehat{\la_i},\ldots,\widehat{\la_j},\cdots,\la_n)
\mathbb{R}_{n-2}^{(i',j')}(\la_1,\ldots,\widehat{\la_k},
\ldots,\widehat{\la_l},\ldots,\la_n)^{-1},\nn
\en
where $k''$ and $l''$ are the positions of $\la_k$ and $\la_l$
in $(\la_i,\la_j,\la_1,\ldots,\widehat{\la_i},
\ldots,\widehat{\la_j},\ldots,\la_n)$.
Using \eqref{eq:X-red}, 
we have
\be
&&{}_{n-2}\Pi _{n}\circ {}_n\Omega_{n-2}^{(k'',l'')}(\la_i,\la_j,\la _1,
\ldots,\widehat{\la_i},\ldots,\widehat{\la_j},\cdots,\la_n)\\
&&=\begin{cases}
0&\hbox{ if }\{i,j\}\cap\{k,l\}\not=\emptyset;\\
{}_{n-2}\Omega_{n-4}^{(k',l')}(\la_1,\ldots,\widehat{\la_i},
\ldots,\widehat{\la_j},\ldots,\la_n)
\circ {}_{n-4}\Pi _{n-2}
&\hbox{ if }\{i,j\}\cap\{k,l\}=\emptyset.
\end{cases}
\en
Combining these two equalities, we obtain the assertion.
\end{proof}

The exchange relation for $\widehat{\Omega}_n^{(i,j)}$ 
takes the following form. 
\begin{lem}
The operators 
$\widehat{\Omega}_n^{(i,j)}$ satisfy the exchange relations:
\bea
&&
\Rc_{k,k+1}(\la_{k,k+1})
\Rc_{\overline{k+1},\overline{k}}(\la_{k+1,k})
\widehat{\Omega}_n^{(i,j)}(\cdots,\la_{k},\la_{k+1},\cdots)=
\label{eq:Om-sym}
\\&&=
\widehat{\Omega}_n^{(\pi_k(i),\pi_k(j))}
(\cdots,\la_{k+1},\la_k,\cdots)
\Rc_{k,k+1}(\la_{k,k+1})
\Rc_{\overline{k+1},\overline{k}}(\la_{k+1,k}),
\nn
\ena
where $\pi_k$ denotes the transposition $(k,k+1)$. 
\end{lem}
\begin{proof}
The statement follows immediately
from \eqref{eq:X-Rsym}
and the relation
$$
{}_{n-1}\Pi_n\circ
\Rc_{1,2}(\la_{l,l'})
\Rc_{\overline{2},\overline{1}}(\la_{l',l})
={}_{n-1}\Pi_n. 
$$
\end{proof}

An important property of 
$\widehat{\Omega}_n^{(i,j)}$ is their commutativity:
\begin{lem}
The operators $\widehat{\Omega}_n^{(i,j)}$ are commutative,  
\begin{align}
&\widehat{\Omega}_n^{(i,j)}\widehat{\Omega}_n^{(k,l)}
=\widehat{\Omega}_n^{(i,j)}\widehat{\Omega}_n^{(k,l)}
\quad \mbox{for all $\quad$ $i<j$, $k<l$}. 
\label{comm2}
\end{align}
\end{lem}
\begin{proof}
In view of the exchange relation \eqref{eq:Om-sym}, 
the proof is reduced to the case $(i,j)=(1,2)$, $(k,l)=(3,4)$. 
Using Lemma \ref{lem:OmOm} and noting that 
$\mathbb{R}_n^{(1,2)}=1$, we obtain 
\be
&&\widehat{\Omega}_n^{(1,2)}\widehat{\Omega}_n^{(3,4)}
=16\ {}_n\Omega^{(1,2)}_{n-2}(\la_1,\cdots,\la_n)
\circ {}_{n-2}\Omega^{(1,2)}_{n-4}(\la_3,\la_4,\cdots,\la_n)
\circ {}_{n-4}\Pi_{n}\nn\\ 
&&\hspace{6em} \times
\mathbb{R}_n^{(3,4)}(\la_1,\cdots,\la_n)^{-1},
\\
&&
\widehat{\Omega}_n^{(3,4)}
\widehat{\Omega}_n^{(1,2)}
=
16\ {}_n\Omega^{(3,4)}_{n-2}(\la_1,\cdots,\la_n)\circ
{}_{n-2}\,\Omega^{(1,2)}_{n-4}(\la_1,\la_2,\la_5,\cdots,\la_n)
\circ {}_{n-4}\Pi_{n}.
\en
Since 
\be
{}_{n-4}\Pi_{n}\circ \mathbb{R}_n^{(3,4)}(\la_1,\cdots,\la_n)^{-1}
={}_{n-4}\Pi_{n}, 
\en
the Lemma reduces to the commutativity \eqref{eq:X-comm}.
\end{proof}

Using Lemma \ref{lem:OmOm} and the equality \eqref{RRSS},
we see that
\begin{align}
&\widehat{\Omega}_n^{(i_1,j_1)}\widehat{\Omega}_n^{(i_2,j_2)}\cdots
\widehat{\Omega}_n^{(i_m,j_m)}\bs_{n}=\label{ooo}
\\&=\begin{cases}
(-4)^m\Omega_{K_1,(i_1,j_1)}\circ
\cdots \circ\Omega_{K_{m},(i_m,j_m)}
\left(\bs_{n-2m}\right)\ &\hbox{if }\{i_1,j_1,\ldots,i_m,j_m\}\hbox{ are distinct};\\
\qquad 0
&\hbox{otherwise.}
\end{cases}\nn
\end{align}
In comparison to the original formula, 
this provides a certain progress since in the left hand side the 
number of $\la$'s does not change along the product. 

Consider the operator 
\bea
&&\widehat{\Omega}_n 
=
\sum\limits _{i<j}\widehat{\Omega}^{(i,j)}_n.
\nn
\label{eq:Ohat}
\ena
{}From the formulae \eqref{ooo} we find the following nice  
representation for $h_n$:
\begin{thm}\label{thm:e^O}
The formula for $h_n(\la _1, \cdots ,\lambda _n)$ 
can be rewritten  
using the operator $\widehat{\Omega _n}$:
\begin{align}
h_n(\la _1, \cdots ,\lambda _n)=2^{-n}
e^{\widehat{\Omega}_n(\la _1, \cdots ,\lambda _n)}\ \bs_{n}.
\label{newform}
\end{align}
\end{thm}
\begin{proof}
Expand the exponential 
into a power series in $\widehat{\Omega}_n$. 
Due to commutativity \eqref{comm2} and the first relation in Lemma \ref{lem:OmOm}, 
we have the nilpotent property 
$$
\frac 1 {m!}(\widehat{\Omega}_n)^m=0\quad \text{for} \quad  
m>\left[\frac n2\right],  
$$
since for such $m$ 
there are necessarily coincident indices $(i,j)$ of  
$\widehat{\Omega}_n^{(i,j)}$.
Hence the series terminates.   
The terms
$$\frac 1 {m!}(\widehat{\Omega}_n)^m\quad \text{for} \quad  
m\le\left[\frac n2\right]$$
contain non-vanishing expressions,  
which 
due to the second relation in (\ref{ooo}) can be written as
$$
\Omega_{K_1,(i_1,j_1)}\circ
\Omega_{K_2,(i_2,j_2)}\circ
\cdots \circ\Omega_{K_{m},(i_m,j_m)}
\left( \bs_{n-2m}\right)
$$
with 
$\{i_p,j_p\}\cap \{i_q,j_q\}=\emptyset$ for all $p,q$. 
For every ordered set $i_1<\cdots <i_m$ there are $m!$ terms, 
all of which are equal due to commutativity \eqref{comm2}. 
\end{proof}

This is a proper place to discuss the problem of 
taking the homogeneous limit $\la _j\to 0$ ($\forall j$). 
At present we do not know how to take this limit,  
but we hope that the formula \eqref{newform} 
will help to solve this problem. 
At least it makes clear that the singularities 
at $\la _k=\la _j$ which persist 
in our formulae are actually spurious. 

\begin{lem}
The operator $\widehat{\Omega}_n(\la _1,\cdots \la _n)$ 
is regular at $\la _k=\la _j$. 
\end{lem}
\begin{proof}
Because of the exchange relation \eqref{eq:Om-sym}, 
it suffices to consider the pole at $\la_1=\la_2$. 
Let us examine each summand which enter the definition of 
$\widehat{\Omega}_n(\la _1,\cdots \la _n)$. 
 
Terms $\widehat{\Omega}^{(i,j)}_n$ with $3\le i<j$ 
do not have a pole. 
The term $\widehat{\Omega}^{(1,2)}_n$ is also regular 
by Lemma \ref{lem:pole}. 
The remaining terms are $\widehat{\Omega}^{(1,i)}_n$ 
and $\widehat{\Omega}^{(2,i)}_n$ with $3\le i$.  
Noting the relation
\be
&&\Rc_{1,2}(\la_{1,2})\Rc_{\bar{2},\bar{1}}(\la_{2,1})
\widehat{\Omega}^{(1,i)}_n(\la_1,\la_2,\cdots) 
\\
&&\quad
=\widehat{\Omega}^{(2,i)}_n(\la_2,\la_1,\cdots)
\Rc_{1,2}(\la_{1,2})\Rc_{\bar{2},\bar{1}}(\la_{2,1})
\en 
and that $\Rc(0)=1$, we see that the 
residues of $\widehat{\Omega}^{(1,i)}_n$ and 
$\widehat{\Omega}^{(2,i)}_n$ 
pairwise cancel.  
\end{proof}

The problem of writing 
down a simple formula for $\widehat{\Omega}_n$ 
in the homogeneous limit remains open. 
Simple idea of presenting $\widehat{\Omega}_n^{(i,j)}$ 
as a specialization of some commuting family of 
transfer matrices with arbitrary dimension and 
spectral parameter 
cannot be true for the following reason. 
The operators $\widehat{\Omega}_n^{(i,j)}$ 
are commutative, but they are nilpotent. 
So, they are not diagonalizable while transfer matrices are. 
We leave the study of these remarkable operators 
for the future. 

\section{Correlation functions of the XXZ model}
\label{sec:Correlation}

In this section we discuss the connection 
between the solutions constructed in Theorem 
\ref{thm:main} and correlation functions of the XXZ spin chain 
with the Hamiltonian 
\bea
H_{XXZ}=\frac{1}{2}\sum_{j}
\left(\sigma^x_j\sigma^x_{j+1}
+\sigma^y_j\sigma^y_{j+1}
+\Delta\,\sigma^z_j\sigma^z_{j+1}\right). 
\label{eq:XXZ}
\ena
Here $\Delta=\cos\pi\nu$.  
We consider the two regimes,
\be
&\nu \in i\R_{>0}& (\mbox{massive regime}), \\
&0<\nu<1         & (\mbox{massless regime}). 
\en
We consider also an inhomogeneous model in which 
a spectral parameter $\la_j$ is associated with each site $j$. 
Let $(E_{\epsilon,\overline{\epsilon}})_j$ be the 
matrix unit acting on the $j$-th site of the lattice. 
Denoting by $|\vac\rangle$ the normalized ground state of 
\eqref{eq:XXZ}, let 
\be
&&\hti_n(\la_1,\cdots,\la_n)
^{\epsilon_1,\cdots,\epsilon_n,
\overline{\epsilon}_n,\cdots,\overline{\epsilon}_1}
=\prod_{p=1}^n(-\eb_p)\,
\langle \vac|
(E_{-\epsilon_1,\overline{\epsilon}_1})_1
\cdots
(E_{-\epsilon_n,\overline{\epsilon}_n})_n
|\vac\rangle 
\en
be the correlation function in the thermodynamic limit. 
It is known \cite{IIJMNT} that the $V^{\otimes 2n}$-valued function
\be
&&\hti_n(\la_1,\cdots,\la_n)
\\
&&=\sum
\hti_n(\la_1,\cdots,\la_n)^{\epsilon_1,\cdots,\epsilon_n,
\overline{\epsilon}_n,\cdots,\overline{\epsilon}_1}
v_{\epsilon_1}\otimes\cdots\otimes v_{\epsilon_n}
\otimes
v_{\overline{\epsilon}_n}\otimes\cdots\otimes 
v_{\overline{\epsilon}_1}
\en
is a solution of the reduced qKZ equation
\footnote{In the massive regime, 
the ground states are two-fold degenerate. 
Matrix elements taken between the different 
ground states satisfy a similar equation with opposite sign.
We do not discuss them here.}
\eqref{eq:Rsymm}, \eqref{eq:rqKZ} as well 
as \eqref{eq:n_to_n-1}. 

{}From the known integral representation \cite{JMMN,JM,KMT},  
it can be shown that $\hti_n$ is meromorphic, 
the only possible poles being simple poles at $\la_i-\la_j=l+l'/\nu$, 
($l,l'\in\Z$, $i\neq j$). Moreover it is holomorphic at $\la_i=\la_j$. 

Let us compare the correlation functions with 
the solution \eqref{eq:def-hn} constructed in Theorem \ref{thm:main}. 
For this purpose we make a specific choice of 
the solutions $\ot(\la)$, $\omega(\la)$ of \eqref{eq:n=2rqKZ}
with similar analyticity properties.  
Set 
\bea
&&\omega(\la):=
c\frac{d}{d\la}
\log\rho(\la)+\frac{1}{4}\frac{[2]}{[\la+1][\la-1]},
\label{eq:om}\\
&&\ot(\la):=
c\frac{d}{d\la}
\log\rt(\la)-\frac{1}{4}\frac{[2\la]}{[\la+1][\la-1]}.
\label{eq:ot}
\ena
where $c=(\sin\pi\nu)/\pi\nu$. 
The functions $\rho(\la)$, $\rt(\la)$ are specified 
in each regime as follows: 
\bea
&&\rho(\la):=
\begin{cases}
\ds{-\zeta\frac{(\zeta^{-2})_\infty(q^2\zeta^{2})_\infty}
{(\zeta^{2})_\infty(q^2\zeta^{-2})_\infty}}
& \mbox{(massive regime)},\\
\ds{-\frac{S_2(-\la)S_2(1+\la)}{S_2(\la)S_2(1-\la)}}
& \mbox{(massless regime)},\\
\end{cases}
\label{eq:rho}
\\
&&\rt(\la):=
\begin{cases}
\ds{\frac{\{q^2\zeta^2\} \{q^6\zeta^{2}\}}{\{q^4\zeta^{2}\}^2}
\frac{\{q^2\zeta^{-2}\} \{q^6\zeta^{-2}\}}{\{q^4\zeta^{-2}\}^2}}
& \mbox{(massive regime)},\\
\ds{\frac{S_3(1-\la)S_3(3-\la)}{S_3(2-\la)^2}
\frac{S_3(1+\la)S_3(3+\la)}{S_3(2+\la)^2}}
& \mbox{(massless regime)}.\\
\end{cases}
\label{eq:rt}
\ena
We have set $\z=e^{\pi i\nu\la}$, 
\be
&&(z)_\infty=(z;q^4)_\infty,\quad \{z\}:=(z;q^4,q^4)_\infty,
\\
&&
S_2(\la):=S_2\bigl(\la|2,\frac{1}{\nu}\bigr),
\quad
S_3(\la):=S_3\bigl(\la|2,2,\frac{1}{\nu}\bigr),
\en
where
$(z;p_1,\cdots,p_r)_\infty
:=\prod_{j_1,\cdots,j_r\ge 0}(1-p_1^{j_1}\cdots p_r^{j_r}z)$,
and $S_r(\la|\omega_1,\cdots,\omega_r)$ denotes the 
multiple sine function. 

\begin{lem}
The functions $\ot(\la),\omega(\la)$ given by \eqref{eq:om}--\eqref{eq:rt}
are the unique solutions of the difference equation \eqref{eq:n=2rqKZ} 
which are holomorphic and bounded in the strip $-1+\delta\le\Re \la\le1-\delta$
for any $0<\delta<1$. 
\end{lem}
\begin{proof}
Direct verification shows that the functions 
\eqref{eq:om}--\eqref{eq:rt} possess the required properties. 
To see the uniqueness, 
suppose $\ot_i(\la),\omega_i(\la)$ are two solutions 
with the same properties, and set $\ot'(\la)=\ot_1(\la)-\ot_2(\la)$, 
$\omega'(\la)=\omega_1(\la)-\omega_2(\la)$. 
Then $\omega'(\la-1)+\omega'(\la)=0$ and  
$\ot'(\la-1)+\ot'(\la)+\omega'(\la)=0$.  
Therefore $\omega'(\la)$ is holomorphic everywhere. 
Together with the boundedness we conclude that $\omega'(\la)=0$. 
Now the same argument applies to show that $\ot'(\la)=0$.  
\end{proof}

Denote by $h^{Ansatz}_n$ the solution 
constructed in Theorem \ref{thm:main} with the above 
choice of $\ot(\la),\omega(\la)$.  
The following lemma shows that the two families of solutions 
$\left\{h_n^{Ansatz}\right\}_{n=0}^\infty$, $\left\{h_n^{Corr}\right\}_{n=0}^\infty$ 
share the same inductive pole structure at $\la_i-\la_j\in\Z\backslash\{0\}$. 

\begin{lem}\label{lem:PPh}
Let $\left\{h_n\right\}_{n=0}^\infty$ be a solution of the system 
\eqref{eq:Rsymm}--\eqref{eq:n_to_n-1}. 
Assume further that $h_n$ is holomorphic at $\la_i=\la_j$  
for all $i\neq j$. Then it is also holomorphic at $\la_i=\la_j\pm 1$. 
We have 
\bea
&&\cP^-_{1,2}h_n(\la_2-1,\la_2,\cdots,\la_n)
=\frac{1}{2}s_{1,2}s_{\bar 1,\bar2}
h_{n-2}(\la_3,\cdots,\la_n)_{3,\cdots,n,\bar{n},\cdots,\bar{3}}, 
\label{eq:P12h}
\\
&&c^{-1}\res_{\la_i=\la_j+l}h_n(\la_1,\cdots,\la_n)
\label{eq:res_h}
\\
&&\quad
=(-1)^{l-1}{}_{n}X_{n-2}^{(i,j)}(\la_1,\cdots,\la_n)\Bigl|_{\la_{i,j}=l}
\left(h_{n-2}(\la_1,\cdots,\widehat{\la_i},\cdots,
\widehat{\la_j},\cdots,\la_n)\right), 
\nn
\ena
where $i<j$, 
$l\in\Z\backslash\{0\}$.  
\end{lem}

\begin{proof}
Since $A^{(k)}_n(\la_1,\cdots,\la_n)$ is holomorphic at 
$\la_i=\la_j$, the regularity of $h_n$ at $\la_i=\la_j\pm 1$ 
is a consequence of the reduced qKZ equation. 
To show \eqref{eq:P12h}, we use the formula
\be
A_n^{(1)}(\la_1,\cdots,\la_n)\Bigl|_{\la_1=\la_2}
&=&(-1)^{n}r_{\bar1,\bar2}(-1)t^{[1,2]}_{n,\bar1}(\la_2)P_{1,\bar1}P_{1,2}
\\
&=&-2P_{1,\bar1}P_{1,2}\cP^-_{2,\bar2} 
A^{(1)}_{n-1}(\la_2,\cdots,\la_n)_{2,\cdots,n,\bar{n},\cdots,\bar{2}}.
\en
Specializing \eqref{eq:rqKZ}, we obtain 
\be
\cP^-_{1,2} h_n(\la_2-1,\la_2,\cdots,\la_n)
&=& 
-2P_{1,\bar1}P_{1,2}\cP^-_{1,\bar1}\cP^-_{2,\bar2}
\\
&\times&
A^{(1)}_{n-1}(\la_2,\cdots,\la_n)_{2,\cdots,n,\bar{n},\cdots,\bar{2}}
h_n(\la_2,\la_2,\cdots,\la_n).
\en
Applying \eqref{eq:n_to_n-1} and then 
using the reduced qKZ equation \eqref{eq:rqKZ} 
for $h_{n-1}$, we rewrite the right hand side as 
\be
&&-P_{1,\bar1}P_{1,2}\cP^-_{2,\bar2}
A^{(1)}_{n-1}(\la_2,\cdots,\la_n)
_{2,\cdots,n,\bar{n},\cdots,\bar{2}}
s_{1,\bar1}h_{n-1}(\la_2,\cdots,\la_n)
_{2,\cdots,n,\bar{n},\cdots,\bar{2}}
\\
&&=-P_{1,\bar1}P_{1,2}\cP^-_{2,\bar2}s_{1,\bar1}
h_{n-1}(\la_2-1,\cdots,\la_n)
_{2,\cdots,n,\bar{n},\cdots,\bar{2}}
\\
&&=-P_{1,\bar1}P_{1,2}\frac{1}{2}s_{1,\bar1}s_{2,\bar2}
h_{n-2}(\la_3,\cdots,\la_n)
_{3,\cdots,n,\bar{n},\cdots,\bar{3}}.
\en
The assertion follows from this.

Let us prove \eqref{eq:res_h}. 
We consider the case $(i,j)=(1,2)$ and $l=-k-1$ with $k\ge 0$. 
The other cases follow from \eqref{eq:Rsymm} 
and \eqref{eq:X-Rsym}. 

Using \eqref{eq:rqKZ} and the regularity of $h_n$ at $\la_1=\la_2$,
we see that the left hand side of \eqref{eq:res_h} is regular at
$\la_1=\la_2-1$. By \eqref{eq:Xla=1}, 
${}_{n}X^{(1,2)}_{n-2}(\la_1,\cdots,\la_n)$ is $0$ for $\la_{1,2}=-1$.
Hence \eqref{eq:res_h} holds for $k=0$. 

For $k=1$, we apply \eqref{eq:Xla=2} to the right hand side of
\eqref{eq:res_h}. We use \eqref{eq:rqKZ} and
\be
c^{-1}\res_{\la_1=\la_2-1}A^{(1)}_n(\la_1,\cdots,\la_n)
=2(-1)^{n}\frac{1}{[2]}r_{\bar1,\bar2}(-2)
t^{[1,2]}_{n,\bar 1}(\la_2-1)P_{1,\bar1}\cP^-_{1,2}
\en
to the left hand side of \eqref{eq:res_h}. Then, using \eqref{eq:P12h},
we obtain the equality.

In the general case $k\ge 2$, 
the residue of both sides of \eqref{eq:rqKZ} gives
\be
\res_{\la_1=\la_2-k-1}h_n(\la_1,\cdots,\la_n)
=A_n^{(1)}(\la_2-k,\la_2,\cdots,\la_n)\;
\res_{\la_1=\la_2-k}h_n(\la_1,\cdots,\la_n). 
\en
Using the difference equation \eqref{eq:X-diff1}, 
we obtain \eqref{eq:res_h} by induction. 
\end{proof}

Let us prove that in the massive case the correlation functions are
given by $h_n^{Ansatz}$. 

\begin{lem}\label{lem:fB}
Let $q$ be a complex number with $0<|q|<1$. 
Consider a $q$-difference equation 
\be
f(q^{-1}\z)=B(\z)f(\z)   
\en
where $B(\z)$ is a matrix and $f(\z)$ 
is a vector valued function.  
Assume that $B(\z)$ is holomorphic at $\zeta=\infty$, 
and $f(\z)$ is holomorphic on $R\le |\z|<\infty$. 
Then $f(\z)$ has at most a pole at $\z=\infty$. 
\end{lem}
\begin{proof}
Set $M=\sup_{|\z|\ge R}|B(\z)|$, 
$K_0=\sup_{R\le |\z|\le |q^{-1}R|}|f(\z)|$.  
For any non-negative integer $n$, we have
in the domain $|q^{-n}R|\le|\z|\le |q^{-n-1}R|$ 
\be
|f(\z)|=|B(q\z)\cdots B(q^n\z)f(q^n\z)|\le M^nK_0. 
\en
Hence there exist $K,N>0$ such that 
\be
|f(\z)|\le K|\z|^N\qquad (|\z|\ge R). 
\en
The assertion follows from this. 
\end{proof}
\medskip

\begin{thm}\label{thm:massive}
In the massive regime, we have 
\be
\hti_n(\la_1,\cdots,\la_n)=h^{Ansatz}_n(\la_1,\cdots,\la_n).
\en
\end{thm}
\begin{proof}
First we note that, in the massive regime,
\be
\prod_{p=1}^n\zeta_p^{(\epsilon_p+\bar{\epsilon}_{p})/2}
\times \left(h_n^{Corr}(\la_1,\cdots,\la_n)\right)
^{\epsilon_1,\cdots,\epsilon_n,\bar{\epsilon}_n,\cdots,\bar{\epsilon}_1}
\en
is a single-valued function of $\zeta_j^2$ ($1\le j\le n$)
(see \cite{JMbk}, eq.(9.4)), 
and that $\ot(\la),\omega(\la)$ are also single-valued 
functions of $\zeta^2$.  
From the definition of ${}_n\Omega^{(i,j)}_{n-2}$ and 
\eqref{eq:F-Parity}, $h_n^{Ansatz}$ has the same functional form. 

To prove the theorem, we proceed by induction on $n$. 
The cases $n=0$ and $1$ are obvious. Suppose 
$h_m^{Corr}=h_m^{Ansatz}$ for $m<n$, and 
consider 
the difference $f(\z)=\hti_n-h^{Ansatz}_n$ 
viewed as a function of $\z=\z_1$.  
We have
\bea
&&f(q^{-1}\z)=B(\z)f(\z) 
\label{eq:fB1}
\ena
where $B(\z)=A^{(1)}_n(\la_1,\cdots,\la_n)$.  
Let us show that $f(\z)\equiv 0$. 

By the induction hypothesis, Lemma \ref{lem:PPh} and the remark made above, 
$f(\z)$ is holomorphic on $\C^\times$ and satisfies 
\bea
&&\cP^{-}_{1,\bar 1}f(\z)=0.
\label{eq:fB2}
\ena
The matrix $B(\z)$ is rational in $\z$ 
and has the following behavior at $\z=\infty$:
\be
B(\z)=\left(-q^{S}+O(\z^{-1})\right)P_{1,\bar1},
\en
where
\be
S=\frac{1}{2}\sigma_{\bar 1}^z
\sum_{p=2}^n(\sigma_p^z+\sigma_{\bar p}^z).
\en
By Lemma \eqref{lem:fB}, $f(\z)$ has at most a pole 
at $\z=\infty$. 
Suppose $f(\z)\not\equiv 0$, and let 
$f(\z)=\z^kf_0+O(\z^{k-1})$ ($f_0\neq 0$) 
be the Laurent expansion. 
Then \eqref{eq:fB1}, \eqref{eq:fB2} stipulate that 
\be
q^{-k}f_0=-q^{S}P_{1,\bar1}f_0=-q^{S}f_0. 
\en
Since $S$ has integer eigenvalues and 
$0<|q|<1$, this is a contradiction.
\end{proof}

Unfortunately, the above reasoning does not carry over to the massless regime 
because of the presence of `extra' poles $\la_{i,j}=l+l'/\nu$.  
For $n=2,3$, the equality $\hti_n=h^{Ansatz}_n$ can be shown 
by direct computation of the integrals. 
We have also checked for $n=4$ and some components
that $h_n^{Ansatz}$ agrees with the known results about 
homogeneous chains \cite{KSTS1}. 
On these grounds, we conjecture that $\hti_n=h_n^{Ansatz}$
holds in the massless regime as well. 

Let us comment on the rational limit.
The limit to the XXX model corresponds to replacing $[\mu]$ by
$\mu$ everywhere. 
The functions $\tilde{\omega},\omega$ defined by \eqref{eq:om}--
\eqref{eq:rt} tend in both massive and massless cases 
to the same functions $\tilde{\omega}^{XXX},\omega^{XXX}$
which satisfy the equation (\ref{eq:n=2rqKZ}) and
\footnote{The function $\omega^{XXX}(\la)$ is related to $\omega(\la)$
in \cite{BJMST} by $\omega(\la)=(\la^2-1)\,\omega^{XXX}(\la)$.}
\begin{align}
\tilde{\omega}^{XXX}(\la)= -\la\ \omega ^{XXX}(\la ).\label{oXXX}
\end{align}
The function ${}_{n}X_{n-2}^{(i,j)}$ becomes in this limit
a rational function ${}_{n}X_{n-2}^{(i,j)\ XXX}(\la_1,\cdots ,\la_n)$.
The expressions for  ${}_{n}\Omega_{n-2}^{(i,j)},
\widehat{\Omega}^{(i,j)}_n$ simplify due to (\ref{oXXX}):
\bea
&&
{}_{n}\Omega_{n-2}^{(i,j)\ XXX}(\la_1,\cdots,\la_n)
=\omega ^{XXX}(\la_{i,j})\cdot {}_{n}X_{n-2}^{(i,j)\ XXX}(\la
_1,\cdots,\la_n), 
\nn \\ 
&& 
\widehat{\Omega}^{(i,j)\ XXX}_n(\la _1, \cdots ,\la _n)=
\omega^{XXX}(\la_{ij})\cdot W^{(i,j)\ XXX}_n(\la_1,\cdots,\la _n), 
\label{OXXX}
\ena 
where 
\begin{align}
&W^{(i,j)\ XXX}_n(\la_1,\cdots,\la_n)\label{WXXX}\\&\quad =-4X^{(i,j)\ XXX}_n(\la _1,\cdots,\la _n)
\circ
{}_{n-2}\Pi _{n}
\circ
\mathbb{R}_n^{(i,j)\  XXX}(\la_1,\cdots,\la_n)^{-1}\nn
\end{align}
with $\mathbb{R}_n^{(i,j)\  XXX}$ being given by the same formula
\eqref{eq:complicated} with $R$ replaced by the rational $R$-matrix.

\vspace{0.3cm}

{\it Acknowledgments.}\quad
Research of HB is supported by INTAS grant \#00-00561 and
by the RFFI grant \#04-01-00352.
Research of MJ is partially supported by 
the Grant-in-Aid for Scientific Research B2--16340033.
Research of TM is partially supported by 
the Grant-in-Aid for Scientific Research A1--13304010.
Research of FS is supported by INTAS grant \#03-51-3350 
and by 
EC networks  "EUCLID",
contract number HPRN-CT-2002-00325 and "ENIGMA",
contract number MRTN-CT-2004-5652. 
Research of YT is partially supported by University of Tsukuba Research Project.

HB and FS are grateful for warm hospitality 
to the University of Tokyo where this work was finished.

HB would also like to thank K. Fabricius, F.G{\"o}hmann, 
A.Kl{\"u}mper,  P. Pyatov,
M.Shiroishi and M.Takahashi for useful discussions.

\end{document}